\documentclass[doublecolumn]{aa}
\usepackage{graphicx}
\usepackage{natbib}
\usepackage{graphicx}
\usepackage[a4paper]{hyperref}
\usepackage{txfonts}
\begin{document}
\def\teff{$T\rm_{eff }$}
\def\kms{$\mathrm {km s}^{-1}$}
\def\gtsima
{\hbox{\raise0.5ex\hbox{$>\lower1.06ex\hbox{$\kern-1.07em{\sim}$}$}}}
\def\ltsima
{\hbox{\raise0.5ex\hbox{$<\lower1.06ex\hbox{$\kern-1.07em{\sim}$}$}}}

\title{
The Ital-FLAMES survey of the Sagittarius dwarf Spheroidal galaxy. I.
Chemical abundances of bright RGB stars
\thanks{Based on observations obtained with FLAMES
at VLT Kueyen 8.2m telescope in the program 71.B-0146.}}

   \subtitle{}

\author{
L. \,Monaco \inst{1,2},
M. \,Bellazzini \inst{2},
P. \,Bonifacio \inst{1},
F.R. \,Ferraro \inst{3},
G. \,Marconi \inst{4},
E. \,Pancino \inst{2},
L. \,Sbordone \inst{5,6},
\and S. \,Zaggia \inst{1}
          }

  \offprints{L. Monaco}

\institute{
Istituto Nazionale di Astrofisica --
Osservatorio Astronomico di Trieste, Via Tiepolo 11,
I-34131 Trieste, Italy
\and
Istituto Nazionale di Astrofisica --
Osservatorio Astronomico di Bologna, Italy
I-40127 Bologna, Italy
\and
Universit\`a di Bologna --
Dipartimento di Astronomia,
I-40127 Bologna, Italy
\and
European Southern Observatory, Casilla 19001, Santiago, Chile
\and
Istituto Nazionale di Astrofisica --
Osservatorio Astronomico di Roma, Italy, 
Via Frascati~33, 00040 Monteporzio Catone, Roma
\and
Universit\`a Tor Vergata, Roma
}

\authorrunning{Monaco et al.}
\mail{lmonaco@eso.org}

\titlerunning{Bright RGB stars in the Sgr dSph}

\date{Received  / Accepted }

\abstract{

We present iron and $\alpha$ element (Mg, Ca, Ti) abundances for a sample of 15
Red Giant Branch stars belonging to the main body of the Sagittarius dwarf
Spheroidal galaxy. Abundances have been obtained from spectra collected using
the high resolution spectrograph FLAMES-UVES mounted at the VLT. Stars of our
sample have a mean metallicity of  [Fe/H]=-0.41$\pm$0.20 with a metal poor tail
extending to [Fe/H]=-1.52. The $\alpha$ element abundance ratios are slightly
subsolar for metallicities higher than [Fe/H]\gtsima-1, suggesting a slow  star
formation rate. The [$\alpha$/Fe] of stars having [Fe/H]$<$-1 are  compatible to
what observed in Milky Way stars of comparable metallicity.

\keywords{Stars: abundances --
Stars: atmospheres -- Galaxies: abundances -- Galaxies: evolution --
Galaxies: dwarf -- Galaxies: individual: Sgr dSph}
}
\maketitle{}

\begin{table*}
\caption{Coordinates and atmospheric parameters for the program stars}
\label{coord}
\begin{center}
\begin{tabular}{lccrrrrrrrr}
\hline
\\
Star & \multicolumn{2}{c}{$\alpha\;\; (J2000.0)\;\;\; \delta\;\; (J2000.0)\;\;\;$}&
$V$ & $I$ & $(V-I)$ & v$_{helio}$(km/s) &
$\rm T_{eff}^a$ &log g & $\xi$ & [M/H]  \\
\\
\hline
\\
2300127         &  $18\, 55\, 46.703$  & $-30\, 35\, 24.683$  & 16.09 & 14.08 &   2.01  &  +147.2$\pm$0.58 & 3687&	0.72 &  2.0 & $-1.0$\\
2300196         &  $18\, 55\, 30.778$  & $-30\, 28\, 19.635$  & 16.26 & 14.54 &   1.72  &  +148.0$\pm$0.94 & 3908&	0.97 &  2.3 & $-0.5$\\
2300215         &  $18\, 55\, 19.146$  & $-30\, 30\, 27.978$  & 16.52 & 14.76 &   1.76  &  +154.9$\pm$0.79 & 3877&	1.08 &  1.9 & $-0.5$\\
2409744         &  $18\, 54\, 55.854$  & $-30\, 32\, 43.106$  & 16.34 & 14.54 &   1.80  &  +131.7$\pm$0.80 & 3837&	0.97 &  1.8 & $-0.5$\\
3600230         &  $18\, 53\, 45.818$  & $-30\, 25\, 49.419$  & 16.43 & 14.74 &   1.69  &  +153.8$\pm$0.56 & 3947&	1.08 &  1.6 & $-0.5$\\
3600262         &  $18\, 53\, 22.340$  & $-30\, 23\, 47.172$  & 16.63 & 14.88 &   1.75  &  +156.0$\pm$0.61 & 3882&	1.16 &  1.9 & $-0.5$\\
3600302         &  $18\, 53\, 45.209$  & $-30\, 30\, 55.702$  & 16.65 & 14.86 &   1.79  &  +143.8$\pm$0.67 & 3848&	1.15 &  1.6 & $-0.5$\\
3800199         &  $18\, 55\, 13.453$  & $-30\, 26\, 42.249$  & 15.35 & 13.90 &   1.45  &  +138.8$\pm$0.63 & 4245&	0.72 &  1.9 & $-1.0$\\
3800204         &  $18\, 55\, 5.7440$  & $-30\, 27\, 56.602$  & 15.81 & 14.25 &   1.56  &  +153.2$\pm$0.61 & 4101&	0.93 &  2.4 & $-1.5$\\
3800318         &  $18\, 54\, 58.264$  & $-30\, 28\, 20.165$  & 16.20 & 14.42 &   1.78  &  +151.8$\pm$0.84 & 3856&	0.90 &  1.9 & $-0.5$\\
3800319         &  $18\, 54\, 58.088$  & $-30\, 28\, 58.481$  & 16.18 & 14.80 &   1.38  &  +141.1$\pm$0.88 & 4364&	1.23 &  1.9 & $-1.5$\\
4303773         &  $18\, 54\, 02.120$  & $-30\, 36\, 21.665$  & 15.97 & 14.23 &   1.74  &  +143.1$\pm$0.63 & 3895&	0.80 &  1.9 & $-0.5$\\
4304445         &  $18\, 53\, 40.606$  & $-30\, 35\, 42.879$  & 16.17 & 14.51 &   1.66  &  +119.9$\pm$0.59 & 3976&	0.95 &  1.7 & $-0.5$\\
4402285         &  $18\, 53\, 19.765$  & $-30\, 37\, 40.099$  & 16.48 & 14.97 &   1.51  &  +159.2$\pm$0.61 & 4156&	1.20 &  1.5 & $-0.5$\\
4408968         &  $18\, 53\, 12.886$  & $-30\, 32\, 03.565$  & 16.67 & 15.07 &   1.60  &  +144.0$\pm$0.68 & 4047&	1.26 &  2.0 & $-0.5$\\
                &                      &                      &       &       &         &                  &     &           &      &       \\
3600073$^\star$ &  $18\, 53\, 56.477$  & $-30\, 27\, 20.337$  & 15.53 & 13.59 &   1.94  &  +156.2$\pm$0.67 & 3731&	0.41 &  2.0 & $-0.5$\\
3700178$^\star$ &  $18\, 54\, 18.068$  & $-30\, 29\, 31.259$  & 16.34 & 14.46 &   1.88  &  +149.1$\pm$0.81 & 3770&	0.92 &  2.3 & $-0.5$\\
3800336$^\star$ &  $18\, 55\, 11.635$  & $-30\, 28\, 00.544$  & 16.21 & 14.29 &   1.92  &  +131.1$\pm$0.74 & 3741&	0.83 &  2.0 & $-0.5$\\
4207953$^\star$ &  $18\, 54\, 14.546$  & $-30\, 32\, 34.502$  & 15.94 & 14.11 &   1.83  &  +129.0$\pm$0.89 & 3815&	0.74 &  1.9 & $-0.5$\\
                &                      &                      &       &       &         &                  &     &           &      &     \\
2300168$^\star$ &  $18\, 55\, 20.010$  & $-30\, 26\, 45.824$  & 16.25 & 13.68 &   2.57  &  +136.0$\pm$2.81 & 3599&	     &      &	  \\
3600181$^\star$ &  $18\, 53\, 57.440$  & $-30\, 25\, 09.207$  & 16.09 & 13.90 &   2.19  &  +147.3$\pm$1.23 & 3610&	     &      &	  \\
3700055$^\star$ &  $18\, 54\, 2.6250$  & $-30\, 26\, 48.807$  & 15.57 & 13.45 &   2.12  &  +121.9$\pm$1.34 & 3634&	     &      &	  \\
4207391$^\star$ &  $18\, 54\, 24.849$  & $-30\, 33\, 02.291$  & 15.93 & 13.79 &   2.14  &  +133.7$\pm$2.01 & 3624&	     &      &	  \\
                &                      &                      &       &       &         &                  &     &           &      &     \\
3600127         &  $18\, 53\, 22.441$  & $-30\, 23\, 59.172$  & 15.94 & 14.42 &   1.52  &  -127.6$\pm$0.51 & 4152&	     &      &	  \\	   
%4207391        &  $18\, 54\, 24.849$  & $-30\, 33\, 02.291$  & 15.93 & 2.14  &  +133.7$\pm$2.01 & 3624&   0.55 &  2.1 & -0.5\\
\\
\hline
\end{tabular}
\\
\end{center}
\hbox{$^a$ We adopted a reddening of $E(V-I)=0.18$ }
\hbox{$\star$ Star showing TiO molecular bands in the spectra }
\end{table*}

\begin{table*}
\caption{Mean chemical abundances for the program stars. The signal to noise ratio of the coadded 
spectra and the number of lines used are also reported}
\label{abund0}
\begin{center}
\begin{tabular}{lllclclclc}
\hline
\\
Star$^a$ & S/N & A(Fe) & $n $ & A(Mg) & $n$ & A(Ca) & $n$ & A(Ti) & $n$\\
         & @653nm &     &      &       &     &       &	   &	    &	 \\
\hline
\\
%                FeI                   MgI                    CaI                      TiI         
Sun      &    &  7.51          &    &  7.58 	       &   &  6.35	      &    &   4.94 	     &   \\
2300127  & 20 &  6.70$\pm$0.24 & 15 &  6.82 $\pm$0.15  & 2 &  5.35 $\pm$0.10  & 9  &   3.99$\pm$0.06 & 7 \\
2300196  & 20 &  7.02$\pm$0.19 & 15 &  7.03 $\pm$0.18  & 4 &  5.67 $\pm$0.16  & 9  &   4.49$\pm$0.16 & 9 \\
2300215  & 14 &  7.28$\pm$0.18 & 13 &  7.04 $\pm$0.18  & 4 &  5.98 $\pm$0.22  & 9  &   4.84$\pm$0.25 & 9 \\
2409744  & 22 &  7.25$\pm$0.06 & 10 &  7.09 $\pm$0.03  & 2 &  5.82 $\pm$0.08  & 8  &   4.67$\pm$0.13 & 9 \\
3600230  & 21 &  7.34$\pm$0.18 & 16 &  7.24 $\pm$0.17  & 4 &  5.81 $\pm$0.14  & 9  &   4.63$\pm$0.14 & 9 \\
3600262  & 21 &  7.14$\pm$0.18 & 15 &  7.27 $\pm$0.13  & 3 &  5.58 $\pm$0.19  & 9  &   4.34$\pm$0.18 & 9 \\
3600302  & 24 &  7.20$\pm$0.18 & 15 &  7.03 $\pm$0.14  & 4 &  5.85 $\pm$0.11  & 9  &   4.53$\pm$0.17 & 9 \\
3800199  & 32 &  6.41$\pm$0.17 & 15 &  6.52 $\pm$0.08  & 4 &  5.60 $\pm$0.10  & 8  &   4.32$\pm$0.11 & 8 \\
3800204  & 31 &  5.99$\pm$0.08 & 13 &  6.28 $\pm$0.04  & 2 &  4.97 $\pm$0.07  & 5  &   3.65$\pm$0.10 & 8 \\
3800318  & 23 &  6.98$\pm$0.17 & 16 &  6.83 $\pm$0.15  & 4 &  5.88 $\pm$0.18  & 9  &   4.69$\pm$0.17 & 9 \\
3800319  & 21 &  6.14$\pm$0.26 & 32 &  6.56 $\pm$0.09  & 2 &  5.38 $\pm$0.15  & 7  &   4.14$\pm$0.14 & 7 \\
4303773  & 18 &  6.78$\pm$0.15 & 14 &  6.66 $\pm$0.16  & 4 &  5.47 $\pm$0.17  & 9  &   3.98$\pm$0.05 & 7 \\
4304445  & 33 &  7.16$\pm$0.14 & 16 &  7.21 $\pm$0.05  & 2 &  5.76 $\pm$0.21  & 9  &   4.45$\pm$0.13 & 9 \\
4402285  & 22 &  7.22$\pm$0.13 & 15 &  7.11 $\pm$0.17  & 4 &  5.99 $\pm$0.25  & 9  &   4.79$\pm$0.13 & 9 \\
4408968  & 18 &  7.09$\pm$0.16 & 17 &  6.81 $\pm$0.04  & 3 &  5.82 $\pm$0.11  & 9  &   4.56$\pm$0.12 & 9 \\
&&&&&&&&&\\
3600073$^\star$ & 43 &  6.73$\pm$0.18 & 17 &  6.67 $\pm$0.14  & 4 &  5.23 $\pm$0.15  & 9  &   3.87$\pm$0.20 & 9 \\
3700178$^\star$ & 19 &  7.06$\pm$0.19 & 16 &  6.83 $\pm$0.17  & 3 &  5.47 $\pm$0.32  & 9  &   4.16$\pm$0.24 & 9 \\
3800336$^\star$ & 25 &  6.88$\pm$0.16 & 14 &  7.01 $\pm$0.10  & 3 &  5.63 $\pm$0.12  & 8  &   4.40$\pm$0.16 & 9 \\
4207953$^\star$ & 30 &  7.10$\pm$0.13 & 15 &  6.97 $\pm$0.10  & 4 &  5.63 $\pm$0.19  & 9  &   4.51$\pm$0.17 & 8 \\
%              FeI                MgI                  CaI                   TiI         
\\
\hline 
\\
\multispan{3}{$\star$ Star showing TiO molecular bands in the spectra}\\
\multispan{3}{A(X)=log($\frac{X}{H}$)+12.00}
\\
\\
\end{tabular}
\end{center}
\end{table*}

\begin{table}
\caption{Mean abundance ratios for the program star}
\label{abund}
\begin{center}
\begin{tabular}{lrrrr}
\hline
\\
Star      & \multispan1{\hfill[Fe/H]\hfill}&   [Mg/Fe] & [Ca/Fe] & [Ti/Fe]\\ 
\\
\hline
\\
2300127   &-0.81$\pm$0.24 & +0.05 & -0.19 &-0.14\\
2300196   &-0.49$\pm$0.19 & -0.06 & -0.19 &+0.04\\
2300215   &-0.23$\pm$0.18 & -0.31 & -0.14 &+0.13\\
2409744   &-0.26$\pm$0.06 & -0.23 & -0.27 &-0.01\\
3600230   &-0.17$\pm$0.18 & -0.17 & -0.37 &-0.14\\
3600262   &-0.37$\pm$0.18 & +0.06 & -0.40 &-0.23\\
3600302   &-0.31$\pm$0.18 & -0.24 & -0.19 &-0.10\\
3800199   &-1.10$\pm$0.17 & +0.04 & +0.35 &+0.48\\
3800204   &-1.52$\pm$0.08 & +0.22 & +0.14 &+0.23\\
3800318   &-0.53$\pm$0.17 & -0.22 & +0.06 &+0.28\\
3800319   &-1.37$\pm$0.26 & +0.35 & +0.40 &+0.57\\
4303773   &-0.73$\pm$0.15 & -0.19 & -0.15 &-0.23\\
4304445   &-0.35$\pm$0.14 & -0.02 & -0.24 &-0.14\\
4402285   &-0.29$\pm$0.13 & -0.18 & -0.07 &+0.14\\
4408968   &-0.42$\pm$0.16 & -0.35 & -0.11 &+0.04\\
&&&&\\
3600073$^\star$   &-0.78$\pm$0.18 & -0.13 & -0.34 &-0.29\\
3700178$^\star$   &-0.45$\pm$0.19 & -0.30 & -0.43 &-0.33\\
3800336$^\star$   &-0.63$\pm$0.16 & +0.06 & -0.09 &+0.09\\
4207953$^\star$   &-0.41$\pm$0.13 & -0.20 & -0.31 &-0.02\\
\\
\hline
\\
\multispan{3}{$\star$ Star showing TiO molecular bands in the spectra}\\
\multispan{3}{[X/Y]=log($\frac{X}{Y}$)-log($\frac{X}{Y}$)$_\odot$}
\\
\end{tabular}
\\
\end{center}
\end{table}

\begin{table}
\caption{Errors in the abundances of star \# 3800318 due to uncertainties
in the atmospheric parameters}
\label{errors}
\begin{center}
\begin{tabular}{lrrrrrr}
\hline
\\
 & $\Delta$A(Fe) & $\Delta$A(Mg) &$\Delta$A(Ca) & $\Delta$A(Ti) \\
\hline
\\
$\Delta\xi =  \pm 0.2$ \kms &  $^{-0.08}_{+ 0.10}$ & $\mp 0.04$ & 
$\mp 0.12$ & $^{-0.15}_{+ 0.17}$\\
\\
$\rm \Delta T_{eff} = \pm 100$ K &$ ^{-0.04}_{+0.07}$ & $^{-0.02}_{+0.05}$ &
$\pm 0.10$ & $^{+0.13}_{-0.12}$\\ 
\\
$\Delta \log g = \pm 0.50 $ & $^{+0.16}_{-0.14}$ & $^{+0.08}_{-0.07}$ &
$^{-0.04}_{+0.01}$ & $\pm 0.05$
\\
\\
\hline
\\
\end{tabular}
\end{center}
\end{table}

\section{Introduction}

The Local Group (LG) is a heterogeneous environment. Galaxies in the LG show a
variety of characteristics (e.g. mass, morphology, gas content) and are
evolving under different conditions (e.g. in isolation, on strong dynamical
interaction). Thererore, in principle, they could teach us about galaxy
evolution as much as  globular clusters did concerning stellar evolution.
Chemical abundances and abundance ratios are key ingredients to study the star
formation histories of stellar systems. The modern generation of spectrographs 
mounted on 8-10~m class telescopes allows to investigate the chemical
composition and dynamics of bright stars in LG galaxies but only a handful of
stars have been studied so far
\citep{tol1,tol2,s03,B00,boni04,ful,gei,she98,she01}.  

The commonly accepted paradigm \citep{wr} predicts the formation of large
galaxies from the hierarchical assembly of small fragments similar to the LG
dwarf spheroidals (dSphs). In this framework, the comparison between the
chemical composition of the Milky Way (MW) and LG dSph stars is a first local
testbed for the hierarchical merging model. The chemical composition of LG
stars turned out to be remarkably different from that of MW stars of comparable
metallicities. In particular, LG stars show $\alpha$ element abundance ratios
systematically under-abundant with respect to MW stars \citep[see, for
instance,][]{venn,boni04}. The interpretation of this empirical evidence is
controversial. Is the hierarchical merging a minor process in the assembly of
the MW? Or were the fragments from which the MW formed at early times different
from the nowadays recognizable dSphs? The chemical difference between MW and LG
stars may reflect an environmental difference between dwarfs accreted at early
times (galaxies near the bottom of the pre-MW potential well -- dense
environment) and the surviving dwarfs \citep[galaxies far from the bottom of
the pre-MW potential well -- loose environment, but see][]{robertson,bullock}.

The Sagittarius dSph \citep[hereafter Sgr,][]{s1} is a LG galaxy currently
experiencing strong  and disruptive tidal interactions with the MW
\citep{igi,s2,maje03}.  Therefore, it may provide clues on the influence of
dynamical interactions on the chemical evolution of dwarf galaxies.

It is well-known that the complex stellar content of Sgr \citep[see][ and
references therein]{bump,bhbletter,nucleo} is largely dominated by a population
of old-intermediate age stars \citep[$\sim$6~Gyr, see,
e.g.][]{sdgs2,ls00,bump}. However, some concerns have been raised on mean
metallicity estimates obtained for this population from spectroscopic and
photometric works \citep[see, e.g.][]{musk,B00,cole,bump,boni04}.

The paper is devoted to the assessment of the mean chemical properties of the
Sgr dominant population. We present Fe, Mg, Ca and Ti abundances for a
selected  sample of stars  belonging to this population. In a companion paper
(Bonifacio et al., in preparation) we deal with the issue of the Sgr
metallicity distribution.

[Fe/H] and [$\alpha$/Fe] abundances as well as the trends in the [Fe/H] {\it
vs}  [$\alpha$/Fe] plane constrain the chemical evolution which led to the
formation of the  Sgr dominant population \citep[][]{lm03}. Moreover, mean
[Fe/H] and [$\alpha$/Fe] values are key ingredients to derive reliable age
estimates from the color-magnitude diagrams.

The paper is organized as follows. In \S\ref{obs} we describe the target
selection and the obtained data.  In \S\ref{atm} we describe the procedures 
followed to fix the atmospheric parameters and the chemical analysis.  In
\S\ref{prvw}  we compare the results obtained  with previous works and in 
\S\ref{concl} we discuss our findings.

\section{Observations}\label{obs}
\subsection{Target selection, Data and Equivalent Widths}

As part of the guaranteed time awarded to the Ital-FLAMES consortium,  more
than 400 stars were observed in the Sgr dSph \citep[][]{cast,zaggia} from May
the 23th to 27th, 2003, using the FLAMES facility mounted on the VLT
\citep{pasquini}. Details on the  observations are given in \citet{zaggia}.
FLAMES allows to observe 132 targets in one shot using the intermediate-low
resolution spectrograph GIRAFFE, plus 8 additional targets using the red arm of
the high resolution specrograph UVES. In this paper we present the results
obtained from the UVES spectra.

\begin{figure}
\centering
\resizebox{\hsize}{!}{\includegraphics[clip=true]{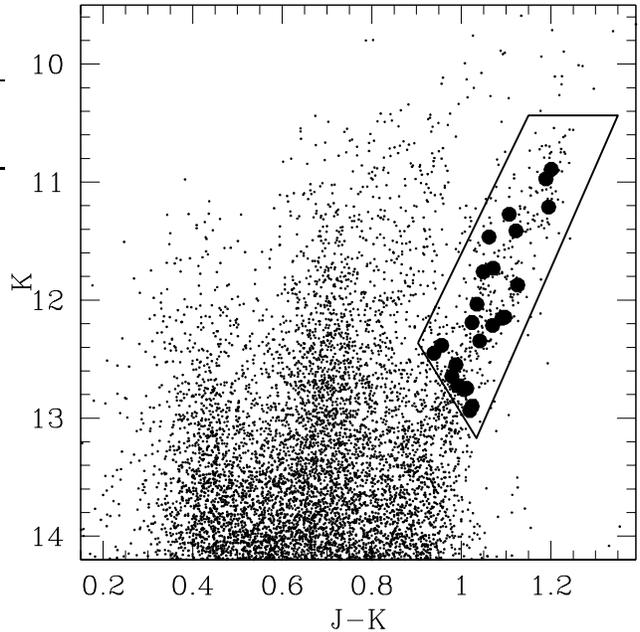}}
\caption{The K {\it vs} J-K$_S$ 2~MASS color--magnitude diagram for a one
square degree region around the globular cluster M~54. Target stars are plotted
as large filled circles.}
\label{cmd2}
\end{figure}

It is important to recall that a large number of Milky Way foreground stars are
present along the Sgr line of sight.  In order to optimize the Sgr star
detection rate, the target selection for the UVES fibres was  performed using
the infrared 2~MASS\footnote{See http://www.ipac.caltech.edu/2mass} color
magnitude diagram (CMD). In fact, in the infrared plane, the upper Sgr red
giant branch (RGB) stands out very clearly from the contaminating MW field
\citep[see, e.g.,][]{cole}.  This also allows a thorough sampling of the Sgr
dominant population. In Fig.~\ref{cmd2} we plotted the 2~MASS (K; J-K$_S$) CMD
for a 1 square degree area centred on the globular cluster M~54. The heavy
continuous line is the selection box. Target stars are plotted as large filled
circles. A similar target selection already proven to be very effective in
detecting stars belonging to the Sgr Stream \citep{maje04}. 

\begin{figure}
\centering
\resizebox{\hsize}{!}{\includegraphics[clip=true]{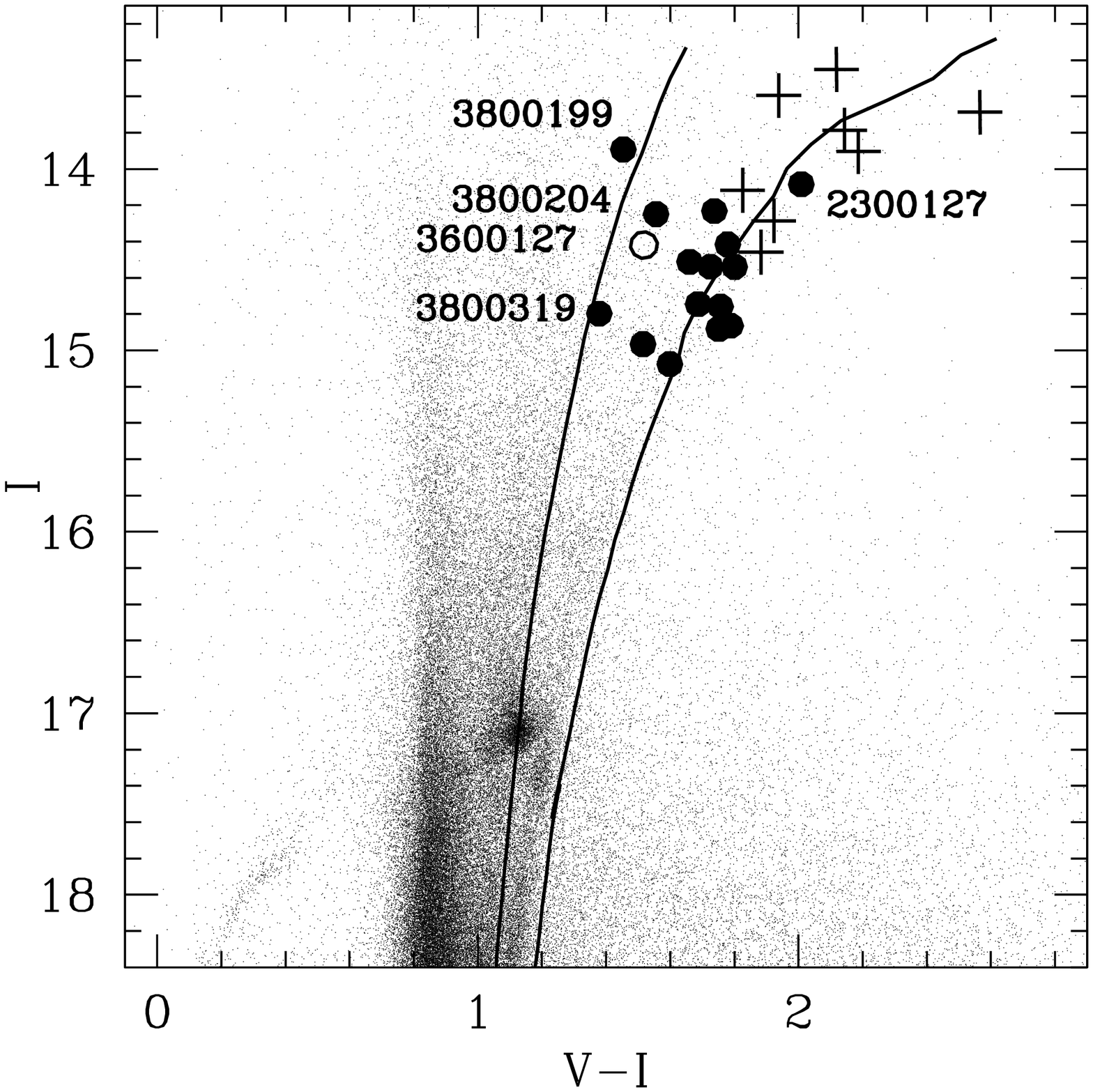}}
\caption{I {\it vs} V-I color--magnitude diagram for a one square degree region
around the globular cluster M~54. Target stars are marked with large symbols. 
Stars showing TiO molecular bands in the spectra are plotted as plus symbols. 
Star 3600217 has a radial velocity not compatible with the membership to Sgr 
and is plotted as a large empty circle. 
Theoretical isochrones from which the surface gravities for the
programme stars were obtained are also plotted as continuous lines.}
\label{cmd}
\end{figure}

Target stars are marked as large symbols in the optical CMD plotted in
Fig.~\ref{cmd} \citep{bump}. In Table~\ref{coord} we report equatorial
(J~2000.0)  coordinates and V and I magnitudes for the target stars.

The coordinates in the J2000.0 absolute astrometric system for both UVES and
GIRAFFE samples were obtained with a procedure already described in other
papers \citep[see, for example,][]{franz01}. The new astrometric Guide Star
Catalogue (GSC II) recently released and now available on the web\footnote{See 
http://www-gsss.stsci.edu/gsc/gsc2/GSC2home.htm }  was used as reference. In
order to derive an astrometric solution we used a program specifically
developed at Bologna Observatory (Montegriffo et al., in preparation). As a
result of the entire procedure, rms residuals of $\sim$0.15 arcsec, both in RA
and Dec, were obtained. The quality of the astrometry was  confirmed by the
successful centreing of the fibres.

We performed the analysis on the spectra reduced  with the  UVES 
ESO-MIDAS\footnote{ESO-MIDAS is the acronym for the European Southern
Observatory Munich Image Data Analysis System which is developed and maintained
by the European Southern Observatory. http://www.eso.org/projects/esomidas/} 
pipeline.  For each pointing, 7 fibres were centred on the target stars while
one fibre was used to measure the sky spectrum. Different spectra of the same
star were  coadded and the resulting signal to noise ratio (S/N) ranges from 14
to 43 at 653~nm (see Table~\ref{abund0}). UVES spectra have a resolution of
R$\simeq$43000  and cover the range between 480~nm and 680~nm.

Equivalent widths (EW) were measured on the spectra using the standard
IRAF\footnote{IRAF is distributed by the National Optical Astronomy
Observatories, which is operated by the association of Universities for
Research in Astronomy, Inc., under contract with the National Science
Foundation.} task {\it splot}. The Fe, Ca, Mg and Ti line lists as well as the
adopted atomic parameters and the measured EW are reported in
Table~\ref{abund1}. A different iron line list (see Table~\ref{abund3800319})
was adopted for star \#3800319  due to the relatively high temperature and
gravity of this star in comparison with the other stars in the sample. We
analysed interactively the spectral lines. For each line the fit has been
visually inspected and adjusted until reaching a satisfying solution.

\subsection{Radial velocities and the contaminating Milky Way field}

Radial velocities (see Table~\ref{coord}) were obtained by cross-correlating
the observed spectra with a rest frame laboratory line list using the recently
released software DAOSPEC\footnote{See
http://cadcwww.hia.nrc.ca/stetson/daospec} (Stetson and Pancino, in
preparation). The final radial velocities and relative errors were  computed
using about 150 lines for each star. Geocentric observed radial velocities were
corrected to heliocentric velocities using the IRAF task {\it rvcorrect}.  

The DAOSPEC code has the capability to measure the line EWs. In our case,
however, we used DAOSPEC only to measure the radial velocities of the target
stars while we used the IRAF task splot to measure EWs for homogeneity with our
previous works on Sgr stars \citep[][]{B00,boni04}. As a check, the
radial velocities of a few stars have also been measured using the {\it fxcor}
IRAF task for Fourier cross correlation. The radial velocities obtained using
DAOSPEC and {\it fxcor} are identical, within the errors. 

\begin{figure}
\centering
\resizebox{\hsize}{!}{\includegraphics[clip=true]{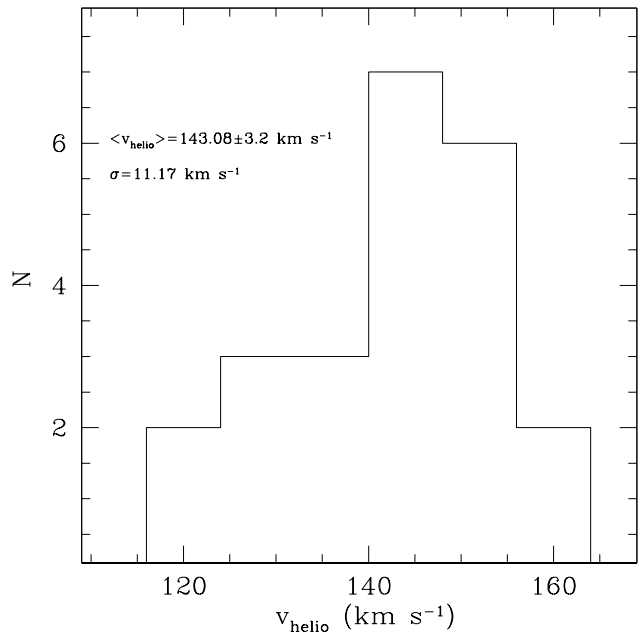}}
\caption{Heliocentric radial velocity distribution for the 23 programme stars
Sgr radial velocity members.}
\label{vr}
\end{figure}

All but one (\#3600127, v$_{helio}$=-127.6~km/s, open circle in Fig.~\ref{cmd})
of the 24 observed stars are indeed Sgr radial velocity members lying within
$\sim$2$\sigma$ of the systemic velocity as measured by \citet{s2}. In
Fig.~\ref{vr} we plotted the velocity distribution of the 23 Sgr radial
velocity members.  The mean velocity
($<$v$_r>$=143.08$\pm$3.2~km~s$^{-1})$\footnote{The quoted 3.2~km~s$^{-1}$ 
error has been estimated employing a bootstrap technique.}  and the velocity
dispersion of the sample ($\sigma$=11.17 ~km~s$^{-1}$) are in good agreement
with the values derived by \citet{igi} and \citet{s2}.  

The MW model of \citet[][ hereafter R03]{r03} predicts that in the M~54 line of
sight 2\% of stars have v$_r>$100~km~s$^{-1}$, if we consider only stars lying
in the  same (V,~V-K)\footnote{The R03 model does not provide the (J-K) color,
therefore we define as selection box in the (V,~V-K) plane the region which
encloses all the target stars.} selection box of the UVES sample.  However, the
model predicts only a $\sim$4\% of giant stars (log~g$<$4)  in the selection
box and none of them with v$_r>$100~km~s$^{-1}$. We checked carefully the 24
stars in the sample and we are confident that all of them are indeed red
giant.  Therefore, even if the R03 model provides only an approximate 
description of the MW,  there is no reason to expect any  MW star among the 23
Sgr radial velocity members in the sample.

\subsection{M-giants showing TiO molecular bands in the spectra}

\begin{figure*}
\centering
\resizebox{\hsize}{!}{\includegraphics[clip=true]{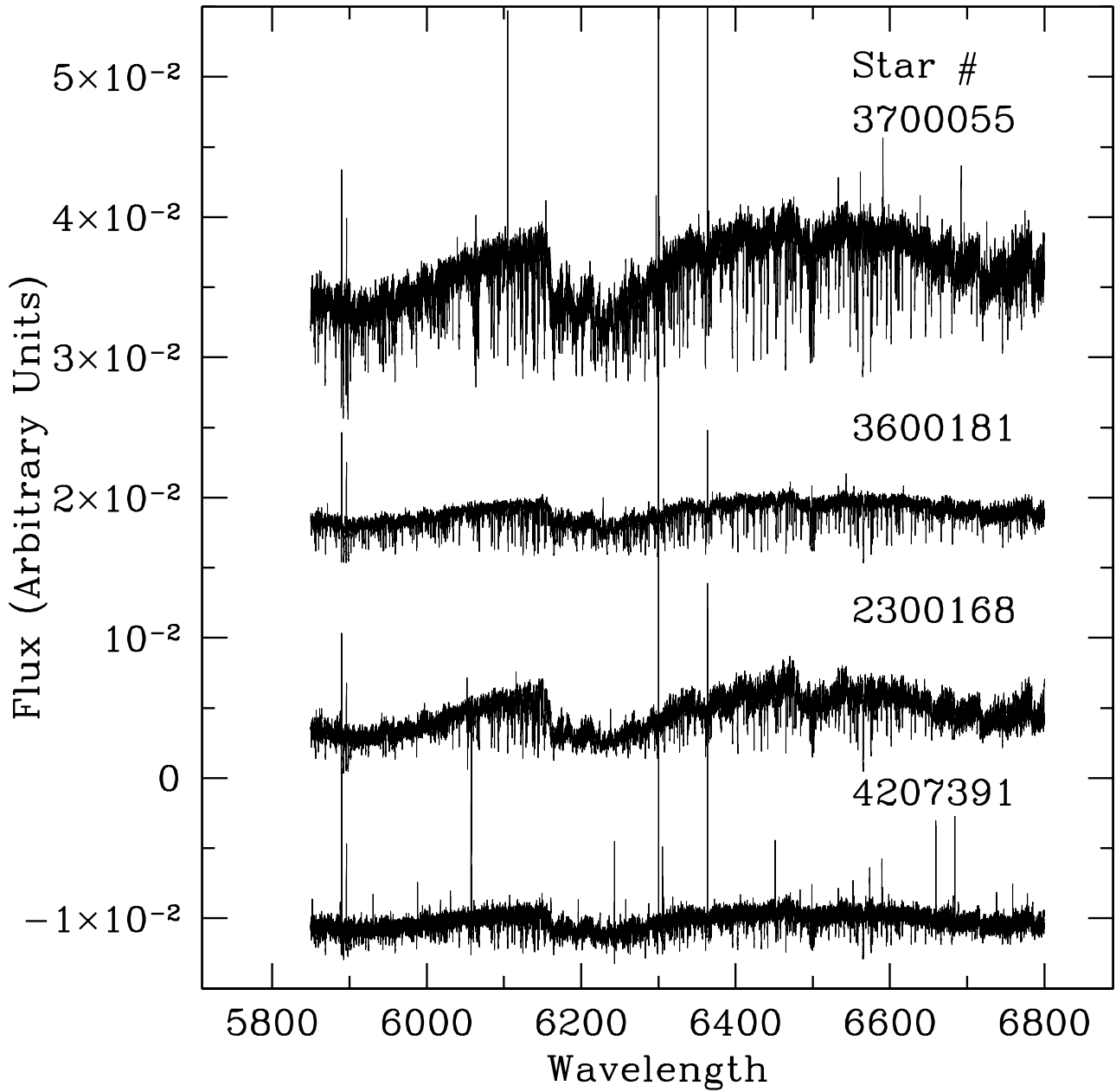}}
\caption{Coadded UVES spectra of the 4 Sgr M-giants having strong TiO bands.}
\label{tio}
\end{figure*}

The coolest (i.e. the reddest) four stars  (\#2300168, \#3600181, \#3700055,
\#4207391, plus symbols at V-I$>$2.0 in Fig.~\ref{cmd}) have effective
temperatures around 3600$^\circ$~K and very strong titanium oxide bands 
\citep[TiO, see][]{tio1,tio2} in the spectra (see Fig.~\ref{tio}). The presence
of the TiO bands confirm these stars as M-giants. Such strong molecular bands
prevent from a safe derivation of the equivalent widths. Therefore, we do not
present the chemical analysis for these stars. In addition, stars \#3600073,
\#3700178, \#3800366, \#4207953 (plus symbols at V-I$<$2.0  in Fig.~\ref{cmd})
show weak but clearly  recognizable TiO bands. For these stars we provide only
a tentative analysis and the derived abundances will not be discussed. We plan
to provide a detailed chemical analysis for these 8  stars by performing
spectral synthesis including also the TiO molecular bands.

In the CMD in Fig.~\ref{cmd} star \#2300127 lies  exactly in the region
occupied by stars with TiO bands in their spectra. Yet this star does not
present any band. The lack of the TiO molecular bands may be due to the
relatively weak Fe and Ti content of this star ([Fe/H]=-0.81, [Ti/Fe]=-0.17,
see Table~\ref{abund}).

\section{Atmospheric Parameters and Chemical analysis}\label{atm}

\begin{figure*}
\centering
\resizebox{\hsize}{!}{\includegraphics[clip=true]{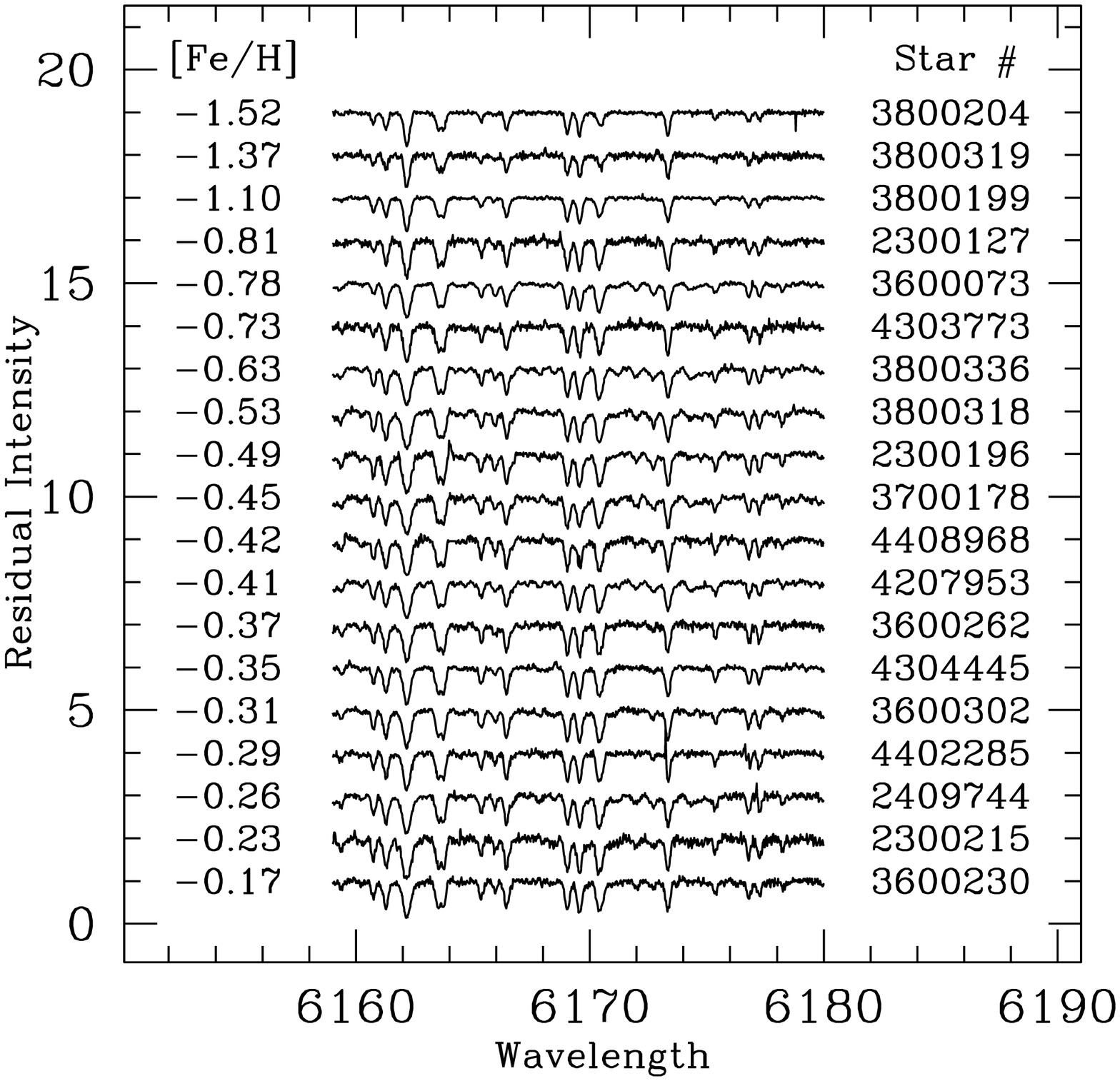}}
\caption{Coadded UVES spectra of the 19 Sgr giants analyzed in this 
paper. Labels on the right denote the star number, those on the
left the [Fe/H].}
\label{data}
\end{figure*}

The UVES spectra of the 19 stars for which the chemical analysis was  performed
(including also stars having weak TiO bands) are plotted in Fig.~\ref{data}. 

\subsection{Effective temperatures and surface gravities}

The effective temperatures for the target stars (see Table~\ref{coord}) were 
derived from   the (V-I) color assuming a reddening of E(V-I)=0.18 
\citep{ls00} and using the calibration of \citet{alonso}. 

We used the \citet{leo} theoretical isochrones, along with E(V-I)=0.18 and 
(m-M)$_0$=17.10 \citep{ls00,tip} as reddening\footnote{We assumed the same
reddening for all the stars in the sample. Inspection of the \citet{dust}
reddening maps provide strong indications that there is no serious variability
of extinction in the considered field \citep[standard deviation of the
reddening value: $\sigma_{E(B-V)}$=0.03,][]{tip}.} and distance modulus, in
order to estimate the gravity of the program stars.  In particular, we used a
(Z=0.001; Age =14.13 Gyr) isochrone for stars \#3800199 \#3800204 \#3800319 and
a  (Z=0.008; Age =6.31 Gyr) isochrone for all the other stars (continuous lines
in Fig.~\ref{cmd}). These two isochrones fit into the range covered by the
target stars on the CMD and the age and metallicity used are also compatible to
what expected from previous works \citep[see][]{bump,ls00,bwg,boni04}.

\subsection{Model atmosphere and Microturbulent velocities}

For each star we computed a plane parallel model atmosphere using version 9 of
the ATLAS code \citep{k93} with the above atmospheric parameters.  Abundances
were derived from EWs using the WIDTH code \citep{k93}.

Microturbulent velocities ($\xi$) were determined minimizing the dependence of
the iron abundance from the EW, among the set of iron lines measured for each
star. 

\begin{figure}
\centering
\resizebox{\hsize}{!}{\includegraphics[clip=true]{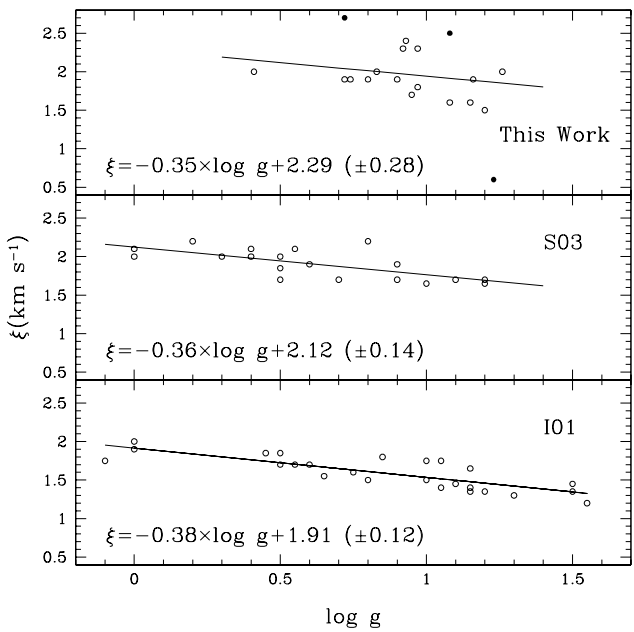}}
\caption{Microturbulent velocities as a function of the surface gravity for the
programme stars (top panel) and stars in the S03 and I01 samples (middle and
bottom panel respectively). Least square fits to the data are plotted as
continuous lines. The filled circles in the upper panel mark stars more than
2-$\sigma$ far from the fitting relation.}
\label{gvt}
\end{figure}

In Fig.~\ref{gvt} we plotted $\xi$ as a function of the adopted gravity for the
stars studied by \citet[][ bottom panel, hereafter I01]{i01}, \citet[][ middle
panel, hereafter S03]{s03} and for stars in our sample (top panel). A clear
trend is present in both the I01 and S03 samples. The same trend is present
also in our sample, albeit with a larger scatter. Continuous lines are least
square fits to the data points. In the case of our sample the fit was obtained
excluding the points having the highest and lowest $\xi$ (2.7 and 0.6
km~s$^{-1}$) and the point with the lowest surface gravity (log g=0.41). As can
be seen, the three fitting lines are very similar to each other.  A weak
dependence of the $\xi$ from the effective temperature was found and it can be
safely neglected as a first order approximation. For stars \#2300127, \#2300215
and \#3800319 (filled circles in fig.~\ref{gvt}) the $\xi$ is 2.7, 2.5 and 0.6
km~s$^{-1}$, respectively, i.e. more than 2-$\sigma$ far from the fitting
relation. When working with low S/N, highly crowded spectra,  it is difficult
to measure weak Fe lines accurately. This may lead to  incorrect $\xi$. Thus,
for stars \#2300127, \#2300215 and \#3800319 we adopted the value obtained from
the  fitting relation $\xi$=-0.35$\times$log~g+2.29, i.e. $\xi$=2.0, 1.9 and
1.9 km~s$^{-1}$, respectively.

\subsection{Chemical Abundances}

The atmospheric parameters (T$_{eff}$, log~g, $\xi$ and the assumed global
metallicity [M/H]) adopted for the program stars are reported in
Table~\ref{coord}. The chemical abundances obtained for each line are reported
in Tables~\ref{abund1} and \ref{abund3800319}. The mean and standard deviation
of such abundances are reported in Tables~\ref{abund0} and \ref{abund} (as
[X/H] abundances in the latter case) for each chemical species. In
Table~\ref{abund0} we also reported the number of lines used to obtain the mean
abundance for each species. The line scatter reported in Table~\ref{abund0}
should be representative of the statistical error arising  from the noise in
the spectra and from uncertainties in the measurement of the  equivalent
widths. 

Under the assumption that each line provides an independent  measure of the
abundance, the error in the mean abundances should be obtained by dividing the
line scatter by $\sqrt{n}$ (where  $n$ is the number of measured lines) and by
adding to this figure the errors arising from the uncertainties in  the
atmospheric parameters. In Table~\ref{errors} we report these latter errors in
the case of star \#3800318, taken as representative of the whole sample.

\begin{figure}
\centering
\resizebox{\hsize}{!}{\includegraphics[clip=true]{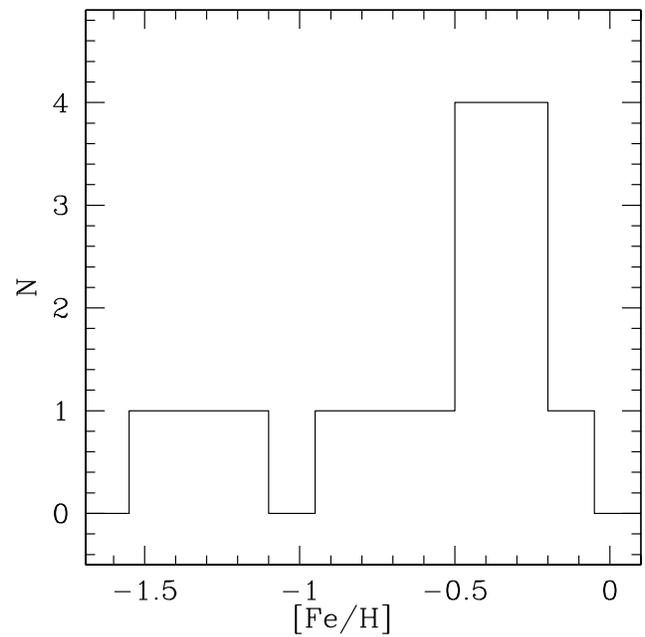}}
\caption{Metallicity distribution of the program stars.}
\label{metdistr}
\end{figure}

In Fig.~\ref{metdistr} we plotted the metallicity distribution obtained.   Our
sample spans a rather large metallicity range (-1.52$\leq$[Fe/H]$\leq$-0.17).
The distribution peaks around [Fe/H]$\simeq$-0.4 and presents an extended metal
poor tail\footnote{Preliminary results obtained from the GIRAFFE sample show
that such tail extends at least down to [Fe/H]$<$-2.5 \citep{zaggia,cast}}.  In
particular, considering only stars more metal rich than [Fe/H]$\simeq$-1, which
should be representative of the Sgr dominant population, we obtain a mean value
of $<$[Fe/H]$>$=-0.41$\pm$0.20.

\begin{figure}
\centering
\resizebox{\hsize}{!}{\includegraphics[clip=true]{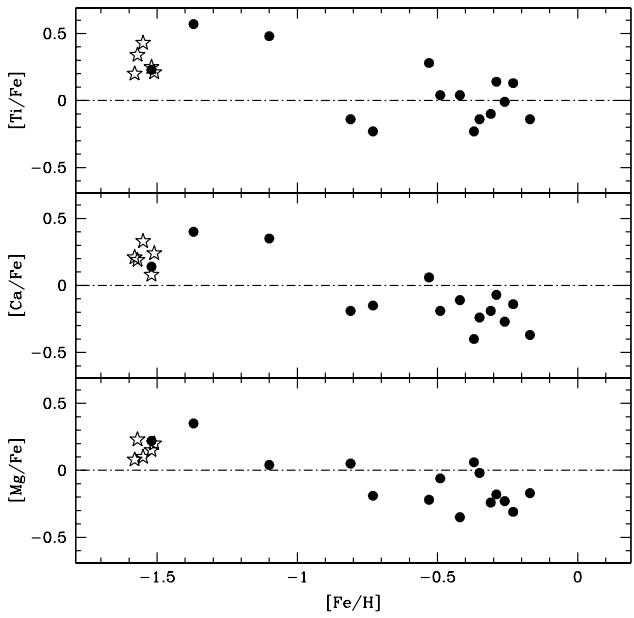}}
\caption{$\alpha$ element abundance ratios ([Ti/Fe], [Ca/Fe] and [Mg/Fe] from 
top to bottom panel) as a function of the iron abundance for the program stars. 
Large open stars mark the 5 M~54 stars studied by B99.}
\label{alpha}
\end{figure}

In Fig.~\ref{alpha} we plotted the [Ti/Fe], [Ca/Fe] and [Mg/Fe] ratios (from
top to bottom panel) for the program stars as a function of the [Fe/H]
abundance.  The 5 M~54 stars studied by \citet[][ hereafter B99]{bwg} are
plotted as large open stars.  Assuming
$<[\alpha/Fe]>=\frac{[Mg/Fe]+[Ca/Fe]}{2}$, we also obtain a mean value of
$<$[$\alpha$/Fe]$>$=-0.17$\pm$0.07 for the  dominant population.

Following \citet[][]{scs93}, these values correspond to a global
metallicity\footnote{The ``global metallicity'' is defined as:
[M/H]=[Fe/H]+log(0.638$\times$10$^{[\alpha/Fe]}$+0.362)} of [M/H]=-0.51, which
is in good agreement with the recent photometric estimate by \citet{bump}.

\subsection{Notes on Metal poor stars: \#3800199, \#3800204,\#3800319}
\label{poor}

The three most metal poor stars (\#3800199 \#3800204 and \#3800319) occupy in
the optical CMD (see Fig.~\ref{cmd}) positions  compatible with the M~54 RGB
(which is roughly represented by the bluer isochrone in the plot). 

The most metal poor star (\#3800204, [Fe/H]=-1.52)  lies very near to the M~54
center ($\sim$1$\arcmin$) and its chemical abundances  (Fig.~\ref{alpha}) are
identical to those of the M~54 stars studied by B99.  Therefore,  it seems
quite likely that this star does indeed belong to M~54. 

Star \#3800319 ([Fe/H]=-1.37) is also quite near ($\sim$1$\arcmin$.4) to the
cluster center but its chemical  composition is only marginally compatible with
M~54 and it will be considered a Sgr field star.  However, we note that
\citet{ls00} claimed a metallicity dispersion of $\sim$0.16~dex for M~54 from
the width of the red giant branch.

Star  \#3800199 ([Fe/H]=-1.10) is placed at 3$\arcmin$.2 from the cluster
center  \citep[which  corresponds to $\sim$7 half light radii,][]{trag} and is
significantly  more metal rich than M~54 ([Fe/H]$\sim$-1.55, B99). Therefore,
we consider star \#3800199 part of the Sgr galaxy field.

\section{Comparison with previous works}\label{prvw}

Beside the present work, chemical abundances have been presented for Sgr RGB
stars by \citet[][ 2 and 10 stars, respectively]{B00,boni04} and by \citet[][
hereafter S02, 14 stars]{mcwilliam}.  \citet[][ hereafter B04]{boni04} also
considered the two stars studied in \citet{B00} obtaining a final sample of 12
stars.

\begin{figure}
\centering
\resizebox{\hsize}{!}{\includegraphics[clip=true]{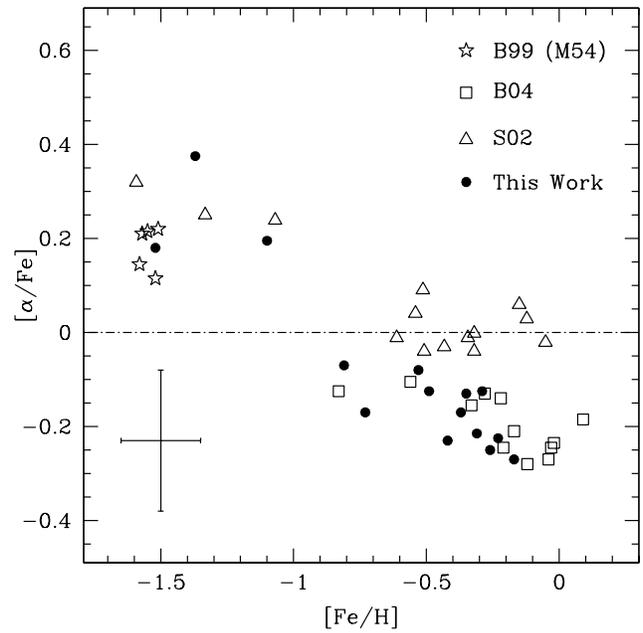}}
\caption{$\alpha$ element abundance ratio --- defined as: 
$[\alpha/Fe]=\frac{[Mg/Fe]+[Ca/Fe]}{2}$ --- as a function of the iron abundance for the 
program stars (filled circles). Large open stars mark the 5 M~54 stars studied 
by B99. Open squares and open triangles mark stars in the B04 and S02
samples, respectively. For stars in the S02 sample: 
$[\alpha/Fe]=\frac{[Si/Fe]+[Ca/Fe]+[Ti/Fe]}{3}$.}
\label{alphamed}
\end{figure}

In Fig.~\ref{alphamed} we plotted the [$\alpha$/Fe] as a function of the iron
abundance for the stars in the 3 samples.  Stars in our sample are plotted as
filled circles, while stars in the B04 and S02 sample are plotted as empty
squares and empty triangles, respectively. The 5 M~54 stars studied by B99 are
marked as large open stars.  The $\alpha$ element abundance ratio is defined as
$[\alpha/Fe]=\frac{[Mg/Fe]+[Ca/Fe]}{2}$ for stars in our sample and in the B04
and B99 samples, while it is defined as
$[\alpha/Fe]=\frac{[Si/Fe]+[Ca/Fe]+[Ti/Fe]}{3}$ for stars in the S02
sample\footnote{S02 do not provide abundances for each species but only mean
values.}.

Stars in the S02 sample range from [Fe/H]$\simeq$-1.6 to [Fe/H]$\simeq$0. In
particular, 3 stars in their sample have  [Fe/H]$<$-1 and 11 stars are in the
range -0.7$\div$0.0. This latter sub-sample has a mean metallicity and $\alpha$
element abundance ratio of: $<$[Fe/H]$>$=-0.36$\pm$0.19 and
$<$[$\alpha$/Fe]$>$=+0.01$\pm$0.04. However, it is important to note that these
values should not be considered as representative of the dominant population,
since their target selection has been biased toward stars with metallicities
within 0.5~dex of the solar value based on previously obtained approximate
metallicities \citep{mcwilliam04}.

The metallicity range of stars in the B04 sample, on the other hand, is
-0.83$\leq$[Fe/H]$<$+0.09. Therefore, it extends toward slightly higher
metallicity with respect to the S02 sample, but it lacks of metal poor stars.
The mean metallicity and $\alpha$ element abundance ratio of the B04 sample
are:  $<$[Fe/H]$>$=-0.23$\pm$0.26 and $<$[$\alpha$/Fe]$>$=-0.20$\pm$0.06.

The mean iron abundance obtained in this paper ([Fe/H]=-0.41) is similar to
that of the S02 and B04 samples.  The 0.18~dex difference between the B04 mean
iron abundance and our figure would be also a little bit lowered (by
0.06$\div$0.09~dex)  by taking into account the different assumption about the 
reddening \citep[B04 adopted E(V-I)=0.22 from][]{marconi}. The different target
selection criterion adopted by B04 may also be responsible for the residual
difference in the mean iron abundance ($\sim$0.1~dex), which is, nevertheless, 
well inside the involved errors.

The $<$[$\alpha$/Fe]$>$ ratio obtained by B04 is very similar to our value 
($<[\alpha/Fe]>=\frac{[Mg/Fe]+[Ca/Fe]}{2}$=-0.17). S02 evaluate the $\alpha$
element abundance ratio as $[\alpha/Fe]=\frac{[Si/Fe]+[Ca/Fe]+[Ti/Fe]}{3}$.
Considering  $[\alpha/Fe]=\frac{[Ca/Fe]+[Ti/Fe]}{2}$, we obtain a 
$<[\alpha/Fe]>$ fairly similar to the S02 figure.  The small residual
difference ($\sim$0.1~dex higher in the S02 sample) may be partly ascribed to
the [Si/Fe] abundances and, possibly, to a different set of lines and atomic
parameters adopted in the chemical analysis. Unfortunately,  S02 neither
provide abundances for each species nor the atomic data and the adopted line
list and this hypothesis cannot be checked further. 

Finally, as already stressed in section~\ref{poor}, we remark that the Fe, Mg,
Ca and Ti abundances of star \#3800204 are consistent with the results obtained
by B99 for M~54 stars.

\section{Discussion and Conclusions}\label{concl}

The main purpose of this paper was to study the chemical composition of the
dominant population of the Sgr dSph galaxy. We selected 24 target stars using
the 2~MASS infrared CMD, where the upper RGB of Sgr is well separated from the
MW field. Target stars have been observed using the red arm of the high
resolution spectrograph FLAMES-UVES.  We reported radial velocities for these
24 stars and all but one are Sgr radial velocity members. Eight stars show
strong or visible TiO bands. For stars with weak TiO bands we present a
tentative chemical analysis while we do not present any chemical analysis for
stars presenting strong TiO bands in the spectra.

For the remaining 15 stars, we reported Fe, Mg, Ca and Ti chemical abundances.
This is the largest sample of high resolution spectra analyzed so far for stars
in the Sgr dSph galaxy, and the only sample thoroughly representative of the
Sgr dominant population.

The metallicity ranges from [Fe/H]=-1.52 to [Fe/H]=-0.17. Three stars have
[Fe/H]$<$-1 and the most metal poor of them (\#3800204) can be reasonably
considered M~54 member.

The mean iron content of stars with [Fe/H]$>$-1 (i.e. the Sgr dominant
population) is $<$[Fe/H]$>$=-0.41$\pm$0.20,  with a mean $\alpha$ element
abundance ratio  $<$[$\alpha$/Fe]$>$=-0.17$\pm$0.07. These figures lead to a
global metallicity [M/H]=-0.51 which is in close agreement with the most recent
photometric estimates obtained for the Sgr dominant population \citep{bump}. 

\begin{figure}
\centering
\resizebox{\hsize}{!}{\includegraphics[clip=true]{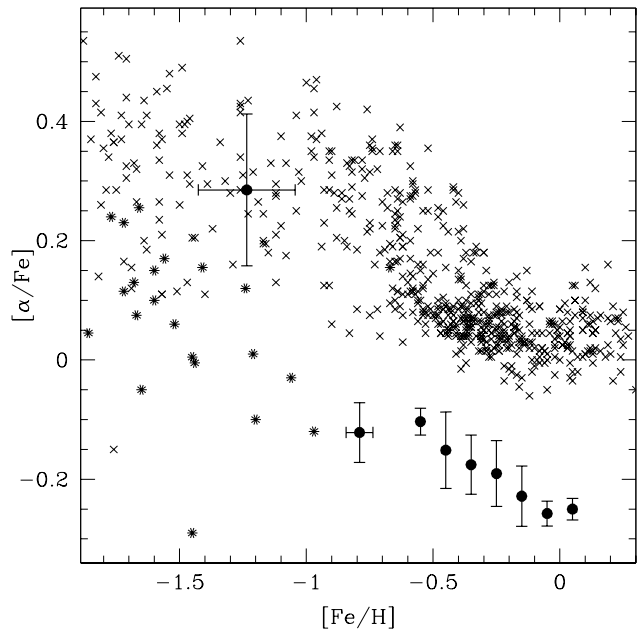}}
\caption{[$\alpha$/Fe]=$\frac{[Mg/Fe]+[Ca/Fe]}{2}$ as a function of [Fe/H] 
for stars in the MW and in Local Group dwarf galaxies 
\citep[crosses and asterisks, respectively, from][]{venn}. 
Filled circles are mean points for Sgr obtained joining
the B04 sample and our data.}
\label{trend}
\end{figure}

In order to obtain a more  statistically significant sample, we now join the
B04 and our samples. In Fig.~\ref{trend} we plotted  in the [$\alpha$/Fe] {\it
vs} [Fe/H] plane the mean points obtained  for Sgr from this larger sample of
Sgr stars as filled circles. For  $-0.65<$[Fe/H]$<0.1$, filled circles
represent running means with 0.20~dex as bin and 0.1~dex as step. For stars
having $-1.0<$[Fe/H]$<-0.65$ and $-1.5<$[Fe/H]$<-1$ (i.e. excluding star
\#3800204 which has been tagged as M~54 member) filled circles are straight
means of the [Fe/H] and [$\alpha$/Fe]  with the corresponding standard
deviations as errorbars. A weak, but  clearly recognizable trend between the
$\alpha$ element abundance ratio and the mean iron abundance exists at high
metallicity. Such a trend waits to be confirmed from a much more extended
sample such as that obtained using the FLAMES-GIRAFFE multifibre spectrograph
which is currently under analysis. For [Fe/H]$<$-1, a sudden increase of the
[$\alpha$/Fe] is apparent.

The mean [$\alpha$/Fe] at low metallicities is consistent with the values
observed in MW stars \citep[crosses in Fig. \ref{trend}, from][]{venn} of
comparable metallicities and somewhat higher with respect to stars in the LG
galaxies \citep[asterisks in the figure, from][]{venn}.  Therefore, metal poor
stars lost in early passages which now are not recognizable as Sgr tidal debris
\citep{helmi}, would be part of the typical content of the MW Halo and
impossible to tag as an accreted component from the chemical composition.

The three metal poor stars in the S02 sample are compatible with MW stars as
well. This occurrence led the authors to suggest that the upper mass end of the
Sgr initial mass function (IMF) should not be significantly different from the
MW one. The level of [$\alpha$/Fe]  which characterizes   a galaxy at low
metallicities may indeed give information on the IMF of the galaxy at that time
\citep[see][ and references therein]{mc}, since the amount of $\alpha$ elements
and iron produced by a Type II SN is a function of the mass of the SN
progenitor.  Although this is true in principle, in practice this information
may not be presently extracted. In fact the ratio of the $\alpha$ elements and
iron produced by a Type II SN is  also a sensitive function of the ``mass
cut'', i.e. the mass coordinate which separates the material of the SN which
``falls back'' on the SN remnant from the material which is ejected. The deeper
the mass cut, the more iron-peak elements are ejected, thus lowering the
overall  [$\alpha$/Fe]. Current SN models are unable to determine the mass cut
in a self consistent way or from first principles: the mass cut is always {\em
assumed}. We do not have either any indication on whether the mass cut is in
any sense ``Universal'' or if it may vary e.g. depending on the mass of the
star or on its metallicity. With this state of affairs, any inference on  the
IMF from the level of the [$\alpha$/Fe] ratio of a galaxy would be highly
uncertain.

The metal-rich Sgr stars lie on the  extension to high metallicity of the
pattern followed by stars in LG galaxies and below MW stars. A low
[$\alpha$/Fe] at high metallicity is traditionally interpreted as evidence for
a slow or bursting star formation rate \citep[S02, B04,][]{gianni}.  On the
contrary, in order to reproduce the Sgr [$\alpha$/Fe] ratios, \citet{lm03}
required a high star formation rate. However, they constrained their model
using preliminary abundances presented by \citet{mcwilliam99}. The somewhat
lower [$\alpha$/Fe] values obtained here and in S02 and B04 should be in better
agreement with a lower star formation rate.

Our data suggest that Sgr had a different chemical evolution from both the MW
and the LG galaxies (see Fig.~\ref{trend}). A different chemical evolution for
Sgr  with respect to the other LG galaxies  is expected, since Sgr experienced
strong and disruptive dynamical interactions with the MW. Such interactions are
witnessed by the Sgr tidal tails studied by \citet{maje03} and are expected to
trigger star formation activity \citep[see, for instance,][]{kra,lucio,zh}.

Finally, we note that the Large Magellanic Cloud (LMC)  metallicity
distribution strongly resambles the Sgr one. In fact, \citet[][]{lmc2}
approximated the metallicity distribution of the LMC bar by two Gaussians
having [Fe/H]=-0.37$\pm$0.15 and [Fe/H]=-1.08$\pm$0.46 and containing 89\% and
11\% of the stars, respectively. The same results hold also for the LMC disk
\citep[see][]{lmc2}.  Clearly, Sgr has the same mean metallicity of the LMC
dominant population as well as the same fraction of metal poor stars
\citep[see][]{bhbletter}. Such  occurrence may suggest a similarity of the Sgr
progenitor with the LMC.

\begin{acknowledgements} 

Part of the data analysis has been performed using software developed by P.
Montegriffo at the INAF - Osservatorio Astronomico di Bologna. This research
was done with support from the Italian MIUR COFIN/PRIN grants 2002028935 and
2004025729. We are grateful to L. Girardi for useful comments and to G.
Schiulaz for a careful reading of the manuscript. 

\end{acknowledgements}

\bibliographystyle{aa}

\appendix

\section{Individual line data} 

The following tables report the  line list and adopted atomic parameters for
the program stars. The measured equivalent width and the corresponding
abundance obtained for each line are also reported.

\begin{table*}
\begin{center}
\caption{Line list and adopted atomic parameters for the program stars. The measured equivalent width and the
corresponding abundance obtained for each line are also reported}
\label{abund1}
{\scriptsize
\begin{tabular}{rrrlrrrrrrrrrr}
\hline
\\
Ion  & $\lambda$ & log gf  & source of   & EW       &$\epsilon$ & EW       & $\epsilon$& EW (pm) & $\epsilon$ & EW      & $\epsilon$ & EW      & $\epsilon$ \\
     & (nm)      &         & log gf      & (pm)     &           & (pm)     &           & (pm)    &            & (pm)    &            & (pm)    &            \\
     &           &         & (see notes) & 2300127  &           & 2300196  &           & 2300215 &            & 2409744 &	     & 3600073$^{\star}$&	     \\
\\
\hline
\\
%%%%%%%%%%%%%%%%%%%%%%%%%%%%%%%%%%%%%%%%%%%%%%%%%%%%%%%%%%%%%%%%%%%%%%%%%%%%%%%%%%%%%%%%%%%%%%%%%%%%%%%%%%%%%%%%%%%%%%%%%%%%%%%%%%%%%%%%%%%%%%%%%%%%%%%%%%%%%%%
Fe I & 585.5076  & -1.76   & FMW          &   --    &	--	&  5.89    & 7.375    &  --      &	--   &  4.42& 7.253         &  4.44     & 7.149    \\ 
Fe I & 588.3817  & -1.36   & FMW          &   9.19  &	6.607	& 10.70    & 6.763    &  11.67   &   7.195   &  --  & --	    & 10.66	& 6.861    \\
Fe I & 595.2718  & -1.44   & FMW          &   6.51  &	6.269	&  9.42    & 6.684    &  --	 &	--   &  --  & --		  &  7.29     & 6.398	 \\
Fe I & 602.4058  & -0.12   & FMW          &  11.52  &	6.654	& 15.93    & 7.153    &  13.97   &   7.173   & 14.75& 7.367	    & 11.14	& 6.569    \\
Fe I & 602.7051  & -1.21   & FMW          &  12.20  &	7.15	& 14.34    & 7.335    &  11.75   &   7.214   &  --  & --	    &  9.30	& 6.637    \\
Fe I & 605.6005  & -0.46   & FMW          &   8.91  &	6.825	&  9.66    & 6.808    &   9.65   &   7.031   &  --  & --	    &  7.34	& 6.535    \\
Fe I & 609.6664  & -1.93   & FMW          &   5.98  &	6.666	& --	   & --       &   9.26   &   7.352   &  --  & --	    &  6.18	& 6.702    \\
Fe I & 615.1617  & -3.30   & FMW          &  15.14  &	6.671	& 17.38    & 6.982    &  17.52   &   7.339   & 16.63& 7.263	    & 15.05	& 6.701    \\
Fe I & 616.5360  & -1.55   & FMW          &   7.46  &	6.772	&  9.78    & 7.066    &   9.20   &   7.19    &  8.86& 7.172	    &  6.80	& 6.657    \\
Fe I & 618.7989  & -1.72   & FMW          &   8.18  &	6.755	& 10.88    & 7.108    &   9.67   &   7.149   & 10.19& 7.291	    &  9.04	& 6.895    \\
Fe I & 622.6734  & -2.22   & FMW          &   2.34  &	6.116	&  7.57    & 7.045    &   7.73   &   7.229   &  7.63& 7.238	    &  5.37	& 6.706    \\
Fe I & 651.8366  & -2.75   & FMW          &  13.18  &	6.811	& 14.61    & 6.979    &  17.20   &   7.677   & 13.50& 7.162	    & 13.83	& 6.937    \\
Fe I & 659.7559  & -1.07   & FMW          &   5.97  &	7.017	&  6.07    & 6.968    &   7.34   &   7.32    &  6.72& 7.248	    &  4.54	& 6.747    \\
Fe I & 670.3566  & -3.16   & FMW          &  11.49  &	6.823	& 13.50    & 7.107    &  14.97   &   7.632   & 12.19& 7.225	 & 11.57     & 6.856	\\
Fe I & 673.9521  & -4.95   & FMW          &  11.75  &	6.714	& 11.36    & 6.785    &  --	 &  --       &  --  & --	    &  9.83	& 6.486    \\
Fe I & 674.6954  & -4.35   & FMW          &  --     &	--      &  6.23    & 7.124    &  --	 &  --       &  6.66& 7.284	    &  5.01	& 6.843    \\
Fe I & 679.3258  & -2.47   & FMW          &   2.41  &	6.658   & --       & --	      &   4.19   &   7.151   &  --  & --		 &  2.73     & 6.742	\\
%%%%%%%%%%%%%%%%                                                                                              
\\
Mg I & 552.8405  & -0.522  & G03          &  23.92  & 	6.67    & 25.91    & 6.811    & 24.36	 & 6.827      & --    & --	 & 20.64     & 6.485	\\
Mg I & 571.1088  & -1.729  & G03          &  14.75  & 	6.98    & 16.63    & 7.125    & 16.58	 & 7.326      & 14.16 & 7.052	 & 12.13     & 6.612	  \\
Mg I & 631.8717  & -1.945  & G03	  &  --	    &   --      &  7.78    & 7.279    &  5.24	 & 7.012      & --    &    --	 &  4.74     & 6.868	\\
Mg I & 631.9237  & -2.165  & G03   	  &  --	    &   --      &  3.67    & 6.913    &  3.73	 & 6.981      &  4.49 & 7.12	 &  2.68     & 6.709 \\
%%%%%%%%%%%%%%%%
\\
Ca I & 585.7451  &  0.240  & SR		  &  19.82  &	5.381	& 23.93   & 5.848    & 21.37	& 5.782      & 20.52 & 5.719   & 17.66  & 5.192   \\
Ca I & 586.7562  & -1.490  & G03		  &   6.94  &	5.263	&  8.57   & 5.643    & 12.08	& 6.253      &  9.27 & 5.811   &  6.25  & 5.231   \\
Ca I & 612.2217  & -0.315  & SR		  &  30.67  &	5.327	& 30.52   & 5.576    & 41.66	& 6.086      &  --   & --      & 26.93  & 5.298   \\
Ca I & 616.9042  & -0.797  & SR		  &  17.36  &	5.494	& 19.67   & 5.846    & 19.24	& 6.089      & 17.68 & 5.908   & 14.76  & 5.151   \\
Ca I & 643.9075  &  0.390  & SR		  &  26.02  &	5.338	& 29.40   & 5.736    & 28.96	& 5.854      & 28.92 & 5.861   & 24.25  & 5.322   \\
Ca I & 645.5558  & -1.290  & SR		  &  14.86  &	5.535	& 15.11   & 5.632    & 14.67	& 5.795      & 13.82 & 5.687   & 13.84  & 5.439        \\
Ca I & 649.3781  & -0.109  & SR		  &  21.22  &	5.289	& 21.69   & 5.351    & 22.97	& 5.812      & 22.43 & 5.795   & 18.13  & 4.933   \\
Ca I & 649.9650  & -0.818  & SR		  &  16.09  &	5.243	& 20.09   & 5.843    & 21.65	& 6.364      & 18.07 & 5.908   & 16.61  & 5.4	  \\
Ca I & 650.8850  & -2.110  & NBS	  &   7.53  &	5.276	&  7.99   & 5.550    &  9.06	& 5.744      &  9.75 & 5.836   &  5.49  & 5.082   \\
%%%%%%%%%%%%%%%%
\\
Ti I & 588.0269  & -2.045  & MFW	  &  13.16  &	4.002	& 14.99   & 4.512    & 16.08	& 4.985      & 14.31 & 4.684   &  9.68  & 3.591    \\
Ti I & 590.3315  & -2.145  & MFW	  &  12.61  &	4.029	& 16.09   & 4.789    & 12.98	& 4.544      & 13.38 & 4.624   & 10.91  & 3.884    \\
Ti I & 593.7809  & -1.890  & MFW	  &  13.52  &	3.903	& 15.31   & 4.403    & 15.86	& 4.761      & 15.38 & 4.705   & 13.95  & 4.088    \\
Ti I & 595.3160  & -0.329  & MFW	  &  15.25  &	4.023	& 17.71   & 4.516    & 17.94	& 4.935      & 17.02 & 4.826   & 11.69  & 3.516    \\
Ti I & 597.8541  & -0.496  & MFW	  &  14.21  &	3.981	& 15.55   & 4.325    & 17.77	& 5.039      & 15.03 & 4.567   & 12.59  & 3.8	   \\
Ti I & 601.6995  & -3.630  & MFW	  &  --     &	   --	&  3.21   & 4.615    &  4.47	& 4.783      &  3.78 & 4.611   &  1.69  & 3.956    \\
Ti I & 606.4626  & -1.944  & MFW          &  --     &	   --	& 13.98   & 4.219    & 16.58	& 4.896      & 16.49 & 4.925   & 13.91  & 4.08     \\
Ti I & 609.1171  & -0.423  & MFW	  &  11.49  &	4.081	& 13.57   & 4.553    & 16.34	& 5.236      & 12.30 & 4.571   & 10.83  & 4.063    \\
Ti I & 609.2792  & -1.379  & MFW	  &   7.80  &	3.89 	& 10.40   & 4.511    &  8.53	& 4.35       &  9.65 & 4.487   &  7.01  & 3.89     \\
%%%%%%%%%%%%%%%%
\\
\hline
\\
\multispan{3}{$\star$ Star showing TiO molecular bands in the spectra}
\\
\\
\multispan{10}{FMW -- Fuhr et al.~(1988)\hfill}\\
\multispan{10}{G03 -- Gratton et al.~(2003)\hfill}\\
\multispan{10}{SR -- Smith et al.~(1981)\hfill}\\
\multispan{10}{NBS -- Wiese et al.~(1969)\hfill}\\
\multispan{10}{MFW -- Martin et al.~(1988)\hfill}\\
\end{tabular}
}
\end{center}
\end{table*}

\addtocounter{table}{-1}

\begin{table*}
\begin{center}
\caption{Line list and adopted atomic parameters for the program stars. The measured equivalent width and the
corresponding abundance obtained for each line are also reported (continued)}
\label{abund1bis}
{\scriptsize
\begin{tabular}{rrrlrrrrrrrrrr}
\hline
\\
Ion  & $\lambda$ & log gf  & source of   & EW       &$\epsilon$ & EW       & $\epsilon$& EW (pm) & $\epsilon$ & EW      & $\epsilon$ & EW      & $\epsilon$ \\
     & (nm)      &         & log gf      & (pm)     &           & (pm)     &           & (pm)    &            & (pm)    &            & (pm)    &            \\
     &           &         & (see notes) & 3600230  &           & 3600262  &           & 3600302 &            & 3700178$^{\star}$&	     & 3800199 &	     \\
\\
\hline
\\
%%%%%%%%%%%%%%%%%%%%%%%%%%%%%%%%%%%%%%%%%%%%%%%%%%%%%%%%%%%%%%%%%%%%%%%%%%%%%%%%%%%%%%%%%%%%%%%%%%%%%%%%%%%%%%%%%%%%%%%%%%%%%%%%%%%%%%%%%%%%%%%%%%%%%%%%%%%%%%%
Fe I & 585.5076  & -1.76   & FMW          &   6.73  & 7.701     &  --  	     &  --      & --       & --	        &  4.42 & 7.217     &  3.06 	& 6.832  \\
Fe I & 588.3817  & -1.36   & FMW          &  11.04  & 7.257  	& 9.78       &6.87  	& 8.95     & 6.902	& 12.59 & 7.134     &  8.77 	& 6.514  \\
Fe I & 595.2718  & -1.44   & FMW          &  10.48  & 7.249  	& 9.60       &6.952 	&10.97     & 7.428	&  --   & --	    &  7.28 	& 6.376  \\
Fe I & 602.4058  & -0.12   & FMW          &  14.35  & 7.397  	&14.91       &7.335 	&11.72     & 7.037	& 12.74 &6.777      &  9.52 	& 6.18   \\
Fe I & 602.7051  & -1.21   & FMW          &  10.62  & 7.166  	&10.34       &6.98  	& 9.77     & 7.077	& 11.18 &6.929      &  7.11 	& 6.235  \\
Fe I & 605.6005  & -0.46   & FMW          &  --	    & --     	& --	     & --       & 9.99     & 7.307	&  9.79 &6.939      &  5.76 	& 6.147  \\
Fe I & 609.6664  & -1.93   & FMW          &   9.35  & 7.493  	& 7.49       &7.071 	& 8.41     & 7.388	&  8.95 &7.174      &  4.23 	& 6.378  \\
Fe I & 615.1617  & -3.30   & FMW          &  15.16  & 7.221  	&15.05       &6.991 	&14.02     & 7.062	& 18.70 &7.176      & 11.90 	& 6.428  \\
Fe I & 616.5360  & -1.55   & FMW          &   9.38  & 7.341  	& 7.81       &6.974 	& --       & --         &  9.53 &7.114      &  5.69 	& 6.435  \\
Fe I & 618.7989  & -1.72   & FMW          &  10.02  & 7.352  	&10.75       &7.36  	& 8.67     & 7.163	&  9.35 &6.959      &  6.20 	& 6.415  \\
Fe I & 622.6734  & -2.22   & FMW          &   7.26  & 7.221  	& 8.91       &7.448 	& 8.27     & 7.492	&  5.19 &6.761      &  3.07 	& 6.321  \\
Fe I & 651.8366  & -2.75   & FMW          &  15.48  & 7.638  	&13.44       &7.125 	&13.62     & 7.39	& 13.99 &6.929      & 10.83 	& 6.586  \\
Fe I & 659.7559  & -1.07   & FMW          &   6.60  & 7.23   	& 5.85       &7.087 	& 5.13     & 7.053	&  7.75 &7.328      &  2.73 	& 6.284  \\
Fe I & 670.3566  & -3.16   & FMW          &  10.61  & 7.083  	&12.58       &7.271 	&11.40     & 7.282	& 15.48 &7.423      &  8.72 	& 6.555  \\
Fe I & 673.9521  & -4.95   & FMW          &  12.39  & 7.381  	&10.84       &6.947 	& 9.69     & 6.914	& 11.79 &6.816      &  7.40 	& 6.47   \\
Fe I & 674.6954  & -4.35   & FMW          &   5.59  & 7.18   	& 7.14       &7.384 	& 5.87     & 7.262	&  5.65 &7.053      & --        & --   	 \\
Fe I & 679.3258  & -2.47   & FMW          &   6.63  & 7.608  	& 4.56       &7.238 	& 4.00     & 7.196	&  4.92 &7.243      & --        & --   	 \\
%%%%%%%%%%%%%%%%                                                                                              
\\
Mg I & 552.8405  & -0.522  & G03          & 26.90  & 7.019    & --   	 & --       & 26.42    & 7.003     & 25.88 & 6.831    & 20.68   & 6.617     \\
Mg I & 571.1088  & -1.729  & G03          & 16.13  & 7.383    & 14.74    & 7.102    & 13.29    & 7.044     & 12.35 & 6.621    & 10.03   & 6.425       \\
Mg I & 631.8717  & -1.945  & G03	  &  5.97  & 7.137    &  7.10    & 7.309    &  3.88    & 6.841     &  5.53 & 7.042    &  2.45   & 6.475     \\
Mg I & 631.9237  & -2.165  & G03   	  &  6.41  & 7.428    &  6.33    & 7.413    &  4.91    & 7.246     & --    &	--    &  1.96   & 6.575  \\
%%%%%%%%%%%%%%%%
\\
Ca I & 585.7451  &  0.240  & SR		  & 19.10  & 5.772  &17.91  	&5.4       & 20.24    & 5.78  	   & 18.14 & 5.09   & 15.44 & 5.501   \\
Ca I & 586.7562  & -1.490  & G03		  &  7.53  & 5.712  & 9.25  	&5.814     &  9.00    & 5.859 	   & 10.79 & 5.788  &  3.81 & 5.442   \\
Ca I & 612.2217  & -0.315  & SR		  & 30.57  & 5.856  &26.93  	&5.432     & 31.74    & 5.737 	   & 41.80 & 5.91   & 24.84 & 5.805   \\
Ca I & 616.9042  & -0.797  & SR		  & 16.63  & 6.011  &17.27  	&5.785     & 17.26    & 6.016 	   & 20.98 & 5.916  & 13.52 & 5.657   \\
Ca I & 643.9075  &  0.390  & SR		  & 24.07  & 5.723  &23.66  	&5.415     & 28.40    & 5.894 	   & 27.02 & 5.399  & 21.83 & 5.641   \\
Ca I & 645.5558  & -1.290  & SR		  & 14.45  & 6.063  &15.14  	&5.88      & 13.90    & 5.875 	   & 14.39 & 5.392  & 10.05 & 5.564   \\
Ca I & 649.3781  & -0.109  & SR		  & 20.98  & 5.653  &19.90  	&5.418     & 20.23    & 5.679 	   & 20.86 & 5.125  & 18.14 & 5.604   \\
Ca I & 649.9650  & -0.818  & SR		  & 16.19  & 5.876  &16.61  	&5.629     & 17.58    & 6.019 	   & 15.73 & 5.102  & 13.18 & 5.564   \\
Ca I & 650.8850  & -2.110  & NBS	  &  7.32  & 5.657  & 6.72  	&5.43      &  8.86    & 5.79  	   &  9.14 & 5.527  & --    & --      \\
%%%%%%%%%%%%%%%%
\\
Ti I & 588.0269  & -2.045  & MFW	  & 12.47   & 4.671  &11.71  	& 4.235     & 11.48    & 4.349     & 13.54 & 4.106 &  8.06 & 4.387   \\
Ti I & 590.3315  & -2.145  & MFW	  & 11.11   & 4.527  &11.98  	& 4.395     & 10.54    & 4.285     & 13.31 & 4.192 &  7.33 & 4.412   \\
Ti I & 593.7809  & -1.890  & MFW	  & 13.78   & 4.754  &12.89  	& 4.274     & 14.18    & 4.709     & 14.58 & 4.102 &  8.04 & 4.24    \\
Ti I & 595.3160  & -0.329  & MFW	  & 15.62   & 4.902  &13.83  	& 4.184     & 13.32    & 4.318     & 12.88 & 3.629 & 11.25 & 4.335   \\
Ti I & 597.8541  & -0.496  & MFW	  & 13.19   & 4.531  &12.35  	& 4.068     & 14.25    & 4.649     & 15.84 & 4.209 &  8.98 & 4.144   \\
Ti I & 601.6995  & -3.630  & MFW	  &  3.09   & 4.703  & 3.59  	& 4.661     &  2.81    & 4.48      &  3.24 & 4.382 &  --   & --      \\
Ti I & 606.4626  & -1.944  & MFW          & 13.25   & 4.655  &14.44  	& 4.532     & 13.73    & 4.628     & 14.89 & 4.147 & 10.04 & 4.502   \\
Ti I & 609.1171  & -0.423  & MFW	  & 10.83   & 4.578  &11.86  	& 4.5	    & 12.48    & 4.783     & 14.67 & 4.556 &  6.40 & 4.262   \\
Ti I & 609.2792  & -1.379  & MFW	  &  7.39   & 4.374  & 7.72  	& 4.253     &  9.62    & 4.606     &  8.87 & 4.141 &  3.56 & 4.275   \\
%%%%%%%%%%%%%%%%
\\
\hline
\\
\multispan{3}{$\star$ Star showing TiO molecular bands in the spectra}
\\
\\
\multispan{10}{FMW -- Fuhr et al.~(1988)\hfill}\\
\multispan{10}{G03 -- Gratton et al.~(2003)\hfill}\\
\multispan{10}{SR -- Smith et al.~(1981)\hfill}\\
\multispan{10}{NBS -- Wiese et al.~(1969)\hfill}\\
\multispan{10}{MFW -- Martin et al.~(1988)\hfill}\\
\end{tabular}
}
\end{center}
\end{table*}

\addtocounter{table}{-1}

\begin{table*}
\begin{center}
\caption{Line list and adopted atomic parameters for the program stars. The measured equivalent width and the
corresponding abundance obtained for each line are also reported (continued)}
\label{abund1ter}
{\scriptsize
\begin{tabular}{rrrlrrrrrrrr}
\hline
\\
Ion  & $\lambda$ & log gf  & source of   & EW       &$\epsilon$ & EW       & $\epsilon$& EW (pm) & $\epsilon$ & EW      & $\epsilon$ \\ %  & EW	   & $\epsilon$
     & (nm)      &         & log gf      & (pm)     &           & (pm)     &           & (pm)    &            & (pm)    &            \\ %  & (pm)    &	       
     &           &         & (see notes) & 3800204  &           & 3800318  &           & 3800319 &            & 3800336$^{\star}$&	     \\ %  & 4207391 &	       
\\
\hline
\\
%%%%%%%%%%%%%%%%%%%%%%%%%%%%%%%%%%%%%%%%%%%%%%%%%%%%%%%%%%%%%%%%%%%%%%%%%%%%%%%%%%%%%%%%%%%%%%%%%%%%%%%%%%%%%%%%%%%%%%%%%%%%%%%%%%%%%%%%%%%%%%%%%%%%%%%%%%%%%%%
Fe I & 585.5076  & -1.76   & FMW          & --    &  --     & 4.61    &  7.241   & --    & --     &  --   & --        \\ %    & --	 & --	  
Fe I & 588.3817  & -1.36   & FMW          & 7.89  & 6.138   & 9.56    &  6.771   & --    & --     &  8.71 & 6.641     \\ %    & --	 & --	  
Fe I & 595.2718  & -1.44   & FMW          & 5.21  & 5.89    & 8.47    &  6.692   & --    & --     &  8.23 & 6.675     \\ %    & --	 & --	  
Fe I & 602.4058  & -0.12   & FMW          & 9.69  & 5.95    &14.80    &  7.274   & --    & --     & 12.01 & 6.835     \\ %    & --	 & --	  
Fe I & 602.7051  & -1.21   & FMW          & 5.84  & 5.867   & 9.99    &  6.858   & --    & --     & 10.99 & 7.049     \\ %    & 10.55	 & 6.953  
Fe I & 605.6005  & -0.46   & FMW          & 5.00  & 5.896   & 9.71    &  7.008   & --    & --     &  7.55 & 6.689     \\ %    & --	 & --	  
Fe I & 609.6664  & -1.93   & FMW          &12.91  & 6.007   & 7.39    &  6.993   & --    & --     &  7.17 & 6.986     \\ %    &  6.93	 & 6.934  
Fe I & 615.1617  & -3.30   & FMW          & 1.13  & 5.933   &15.52    &  6.987   & --    & --     & 16.32 & 7.04      \\ %    & 17.40	 & 7.083  
Fe I & 616.5360  & -1.55   & FMW          & 4.00  & 6.027   & 7.74    &  6.9     & --    & --     &  7.25 & 6.853     \\ %    &  --	 &  --    
Fe I & 618.7989  & -1.72   & FMW          & 4.86  & 6.05    & 9.24    &  7.033   & --    & --     &  7.53 & 6.771     \\ %    &  --	 &  --    
Fe I & 622.6734  & -2.22   & FMW          & 2.25  & 6.014   & 7.37    &  7.128   & --    & --     &  7.11 & 7.112     \\ %    &  --	 &  --    
Fe I & 651.8366  & -2.75   & FMW          & 9.51  & 6.086   &12.81    &  6.947   & --    & --     & 12.48 & 6.862     \\ %    &  --	 &  --    
Fe I & 659.7559  & -1.07   & FMW          & --    & --      & 5.63    &  6.994   & --    & --     &  5.52 & 7.041     \\ %    &  --	 &  --    
Fe I & 670.3566  & -3.16   & FMW          & 6.42  & 6.005   &11.91    &  7.083   & --    & --     &  --   & --        \\ %    & 11.29	 & 6.88   
Fe I & 673.9521  & -4.95   & FMW          & 5.76  & 5.998   & 9.47    &  6.648   & --    & --     & 10.34 & 6.71      \\ %    & --	 & --	  
Fe I & 674.6954  & -4.35   & FMW          &  --   &  --     & 5.72    &  7.096   & --    & --     &  5.85 & 7.103     \\ %    &  6.97	 & 7.198  
Fe I & 679.3258  & -2.47   & FMW          &  --   &  --     &   --    &  --      & --    & --     &   --  &  --       \\ %    &  --	 &  --    
%%%%%%%%%%%%%%%%                           		        		  		    		     	      
\\					   		      					    		     	      
Mg I & 552.8405  & -0.522  & G03          &19.95  &  6.239  & 25.04   &  6.858   & 18.45 & 6.469  & 24.82 & 6.866     \\ %    & 23.74	& 6.767    
Mg I & 571.1088  & -1.729  & G03          &10.90  &  6.313  & 14.70   &  7.06    & 10.81 & 6.653  & 14.66 & 7.101     \\ %    & 15.00	& 7.09  
Mg I & 631.8717  & -1.945  & G03	  &  --	  &    --   &  3.54   &  6.702   &  --	 &  --    &   --  & --        \\ %    & --       & --    
Mg I & 631.9237  & -2.165  & G03   	  &  --	  &    --   &  2.38   &  6.681   &  --	 &  --    &  4.07 & 7.067     \\ %    & --       & --	  

%%%%%%%%%%%%%%%%			   		        	      	   		    	   	     	      
\\					   		        	      	   		    		     	      
Ca I & 585.7451  &  0.240  & SR		  &14.48  & 4.969   & 23.54   &  5.949   & 13.20 & 5.347  & 21.15 & 5.654     \\ %    & 17.16	 & 4.984 
Ca I & 586.7562  & -1.490  & G03		  &   --  & --      & 12.35   &  6.278   &   --	 &  --    &  9.14 & 5.667     \\ %    &  7.05	 & 5.222 
Ca I & 612.2217  & -0.315  & SR		  &   --  & --      & 29.75   &  5.621   & 18.71 & 5.267  &   --  & --        \\ %    & 39.51	 & 5.741 
Ca I & 616.9042  & -0.797  & SR		  &11.15  & 4.938   & 17.28   &  5.77    &  9.69 & 5.237  & 16.82 & 5.6       \\ %    & 16.41	 & 5.273 
Ca I & 643.9075  &  0.390  & SR		  &  --   &--	    & 28.72   &  5.848   & 17.55 & 5.28   & 25.90 & 5.529     \\ %    & 26.12	 & 5.345 
Ca I & 645.5558  & -1.290  & SR		  & 8.64  & 5.092   & 14.54   &  5.749   &  7.43 & 5.356  & 14.79 & 5.691     \\ %    & 12.94	 & 5.158 
Ca I & 649.3781  & -0.109  & SR		  &16.22  & 4.873   & 23.18   &  5.826   & 15.61 & 5.477  & 22.53 & 5.658     \\ %    & 20.88	 & 5.197 
Ca I & 649.9650  & -0.818  & SR		  &11.42  & 4.958   & 19.68   &  6.077   & 12.49 & 5.687  & 18.59 & 5.828     \\ %    & 18.23	 & 5.507 
Ca I & 650.8850  & -2.110  & NBS	  &    -- &   --    &  9.91   &  5.841   &   --	 &  --    &  7.59 & 5.389     \\ %    & 10.26	 & 5.582 
%%%%%%%%%%%%%%%%			   		        	  	    		    	   	     	   		  
\\					   		        	  	    		    		     	   		  
Ti I & 588.0269  & -2.045  & MFW	  & 3.97  & 3.56    & 15.47   &  4.822   &  3.66 & 4.069  & 14.00 & 4.405     \\ %    & 16.80	 & 4.569 
Ti I & 590.3315  & -2.145  & MFW	  & 4.44  & 3.744   & 12.62   &  4.436   &  3.67 & 4.187  & 12.11 & 4.19      \\ %    & 12.78	 & 4.003 
Ti I & 593.7809  & -1.890  & MFW	  & 4.94  & 3.55    & 15.34   &  4.621   &  4.30 & 4.029  & 14.55 & 4.336     \\ %    & 14.93	 & 4.076 
Ti I & 595.3160  & -0.329  & MFW	  & 8.42  & 3.599   & 17.81   &  4.875   &  6.63 & 3.917  & 16.24 & 4.463     \\ %    &  --    &    -- 
Ti I & 597.8541  & -0.496  & MFW	  & 6.43  & 3.515   & 16.98   &  4.849   &  7.17 & 4.14   & 14.47 & 4.261     \\ %    &  --    &    -- 
Ti I & 601.6995  & -3.630  & MFW	  &  --   &  --     &  3.94   &  4.657   &   --	 &  --    &  4.33 & 4.521     \\ %    &  4.29	 & 4.302    
Ti I & 606.4626  & -1.944  & MFW          & 6.83  & 3.78    & 16.48   &  4.829   &  6.06 & 4.293  & 16.14 & 4.627     \\ %    & 16.63	 & 4.364 
Ti I & 609.1171  & -0.423  & MFW	  & 4.41  & 3.744   & 13.42   &  4.706   &  5.51 & 4.346  & 13.62 & 4.626     \\ %    & 15.01	 & 4.592 
Ti I & 609.2792  & -1.379  & MFW	  & 1.94  & 3.711   &  9.20   &  4.403   &  --	 &  --    &  8.93 & 4.214     \\ %    & 11.60	 & 4.375 
%%%%%%%%%%%%%%%%
\\
\hline
\\
\multispan{3}{$\star$ Star showing TiO molecular bands in the spectra}
\\
\\
\multispan{10}{FMW -- Fuhr et al.~(1988)\hfill}\\
\multispan{10}{G03 -- Gratton et al.~(2003)\hfill}\\
\multispan{10}{SR -- Smith et al.~(1981)\hfill}\\
\multispan{10}{NBS -- Wiese et al.~(1969)\hfill}\\
\multispan{10}{MFW -- Martin et al.~(1988)\hfill}\\
\end{tabular}
}
\end{center}
\end{table*}

\addtocounter{table}{-1}

\begin{table*}
\begin{center}
\caption{Line list and adopted atomic parameters for the program stars. The measured equivalent width and the
corresponding abundance obtained for each line are also reported (continued)}
\label{abund1quattro}
{\scriptsize
\begin{tabular}{rrrlrrrrrrrrrr}
\hline
\\
Ion  & $\lambda$ & log gf  & source of   & EW       &$\epsilon$ & EW       & $\epsilon$& EW (pm) & $\epsilon$ & EW      & $\epsilon$ & EW      & $\epsilon$ \\
     & (nm)      &         & log gf      & (pm)     &           & (pm)     &           & (pm)    &            & (pm)    &            & (pm)    &            \\
     &           &         & (see notes) & 4207953$^{\star}$ &           & 4303773  &           & 4304445 &            & 4402285 &	     & 4408968 &	     \\
\\
\hline
\\
%%%%%%%%%%%%%%%%%%%%%%%%%%%%%%%%%%%%%%%%%%%%%%%%%%%%%%%%%%%%%%%%%%%%%%%%%%%%%%%%%%%%%%%%%%%%%%%%%%%%%%%%%%%%%%%%%%%%%%%%%%%%%%%%%%%%%%%%%%%%%%%%%%%%%%%%%%%%%%%
Fe I & 585.5076  & -1.76   & FMW          &  5.63 &  7.408    &  3.06  & 6.896     &  4.90 & 7.28     &  4.63  & 7.275  	&  5.11    & 7.309 	\\
Fe I & 588.3817  & -1.36   & FMW          & 11.23 &  7.057    & 10.20  & 6.836     & 11.99 & 7.326    & --     & --     	& 12.23    & 7.2   	\\
Fe I & 595.2718  & -1.44   & FMW          &  --   &   --      &  8.60  & 6.666     & --    & --       &  9.46  & 7.076  	&  8.97    & 6.75  	\\
Fe I & 602.4058  & -0.12   & FMW          & 12.90 &  6.953    & 13.07  & 6.936     & 13.48 & 7.139    & 13.24  & 7.238  	& 14.22    & 7.087 	\\
Fe I & 602.7051  & -1.21   & FMW          & 11.81 &  7.169    &  --    & --        & 11.50 & 7.219    & 10.76  & 7.235  	& 10.57    & 6.91  	\\
Fe I & 605.6005  & -0.46   & FMW          &  8.70 &  6.822    &  8.52  & 6.741     &  9.16 & 6.936    & --     & --     	& 10.45    & 7.048 	\\
Fe I & 609.6664  & -1.93   & FMW          &  7.32 &  6.959    &  6.15  & 6.739     &  8.17 & 7.164    &  7.72  & 7.191  	&  8.38    & 7.137 	\\
Fe I & 615.1617  & -3.30   & FMW          & 16.99 &  7.166    & 14.48  & 6.781     & 13.99 & 6.905    & 13.75  & 7.143  	& 16.13    & 7.114 	\\
Fe I & 616.5360  & -1.55   & FMW          &  8.92 &  7.082    &  5.83  & 6.531     &  8.76 & 7.113    &  9.79  & 7.447  	&  9.19    & 7.105 	\\
Fe I & 618.7989  & -1.72   & FMW          &  9.47 &  7.052    &  7.88  & 6.751     & 10.19 & 7.27     &  9.90  & 7.366  	&  9.18    & 6.993 	\\
Fe I & 622.6734  & -2.22   & FMW          &  8.24 &  7.249    &  4.73  & 6.647     &  7.65 & 7.205    &  6.40  & 7.078  	&  6.36    & 6.966 	\\
Fe I & 651.8366  & -2.75   & FMW          & 14.04 &  7.11     & 11.96  & 6.765     & 14.87 & 7.422    & 12.52  & 7.254  	& 15.48    & 7.373 	\\
Fe I & 659.7559  & -1.07   & FMW          &  6.24 &  7.087    & --     & --        &  7.29 & 7.264    &  7.17  & 7.293  	&  6.75    & 7.126 	\\
Fe I & 670.3566  & -3.16   & FMW          & 12.82 &  7.193    & 11.45  & 6.965     & 11.33 & 7.101    & 10.31  & 7.13   	& 11.01    & 6.972 	\\
Fe I & 673.9521  & -4.95   & FMW          &  --   &   --      &  9.24  & 6.586     & 10.66 & 6.97     &  9.42  & 7.021  	& 10.28    & 6.888 	\\
Fe I & 674.6954  & -4.35   & FMW          &  5.90 &  7.078    & --     & --        &  6.41 & 7.25     &  4.36  & 7.058  	&  6.19    & 7.253 	\\
Fe I & 679.3258  & -2.47   & FMW          &  4.09 &  7.068    &  4.42  & 7.103     &  3.60 & 6.988    &  5.86  & 7.463  	&  5.15    & 7.278 	\\
%%%%%%%%%%%%%%%%                            				             	    	        			  	     
\\					    				             						  	     
Mg I & 552.8405  & -0.522  & G03          & 26.88 & 6.953     & 21.07  & 6.563     & --    & --       & 24.18  & 6.949  	& 23.99    & 6.79       \\
Mg I & 571.1088  & -1.729  & G03          & 13.37 & 6.858     & 13.48  & 6.856     & 15.72 & 7.268    & 13.61  & 7.108  	& 12.82    & 6.774 	\\
Mg I & 631.8717  & -1.945  & G03	  &  6.26 & 7.135     &  2.48  & 6.456     &  6.43 & 7.161    &  7.58  & 7.389  	& --       &  --   	\\
Mg I & 631.9237  & -2.165  & G03   	  &  3.64 & 6.927     &  2.94  & 6.775     & --    & --       &  3.80  & 6.995  	&  3.20    & 6.862 	\\
%%%%%%%%%%%%%%%%			    			      	             	    	         	        	  	     
\\					    				             	    	         	        	  	     
Ca I & 585.7451  &  0.240  & SR		  & 17.97 & 5.36      & 20.31  & 5.702     & 19.18 & 5.761    & 18.74  & 5.982  	& 20.40    & 5.784 	\\
Ca I & 586.7562  & -1.490  & G03	  &  8.13 & 5.588     &  6.72  & 5.47      &  7.18 & 5.663    &  6.97  & 5.865  	&  8.81    & 5.888 	\\
Ca I & 612.2217  & -0.315  & SR		  & 34.59 & 5.836     & 27.12  & 5.529     & 28.52 & 5.794    & 24.80  & 5.827  	& 30.80    & 5.847 	\\
Ca I & 616.9042  & -0.797  & SR		  & 19.00 & 6.02      & 16.17  & 5.619     & 17.49 & 6.103    & 18.10  & 6.549  	& 17.85    & 5.96  	\\
Ca I & 643.9075  &  0.390  & SR		  & 26.92 & 5.7       & 23.20  & 5.406     & 23.37 & 5.637    & 24.34  & 6.013  	& 25.33    & 5.678 	\\
Ca I & 645.5558  & -1.290  & SR		  & 13.85 & 5.593     & 10.57  & 5.163     & 14.76 & 6.059    & 13.22  & 6.149  	& 14.53    & 5.887 	\\
Ca I & 649.3781  & -0.109  & SR		  & 20.17 & 5.405     & 19.33  & 5.347     & 20.40 & 5.755    & 17.64  & 5.704  	& 21.24    & 5.675 	\\
Ca I & 649.9650  & -0.818  & SR		  & 17.16 & 5.648     & 16.77  & 5.655     & 15.94 & 5.77     & 15.75  & 6.106  	& 18.30    & 5.97  	\\
Ca I & 650.8850  & -2.110  & NBS	  &  8.22 & 5.556     &  5.68  & 5.307     &  5.06 & 5.34     &  5.71  & 5.677  	&  7.25    & 5.676 	\\
%%%%%%%%%%%%%%%%			    			      	             	    	         	        	  	    	   
\\					    				             	    	        	        	  	    	   
Ti I & 588.0269  & -2.045  & MFW	  & 16.14 & 4.879     & 10.28  & 4.003     & 11.33 & 4.419    &  9.77  & 4.594  	& 11.74    & 4.461 	\\
Ti I & 590.3315  & -2.145  & MFW	  & 13.87 & 4.574     &  8.84  & 3.92      & 10.86 & 4.457    & 10.49  & 4.837  	& 12.52    & 4.692 	\\
Ti I & 593.7809  & -1.890  & MFW	  & 14.79 & 4.453     & 10.84  & 3.938     & 11.13 & 4.231    & 11.23  & 4.7    	& 13.69    & 4.597 	\\
Ti I & 595.3160  & -0.329  & MFW	  & 16.82 & 4.627     & 12.91  & 4.008     & 15.24 & 4.731    & 14.01  & 4.95   	& 14.05    & 4.368 	\\
Ti I & 597.8541  & -0.496  & MFW	  & 14.40 & 4.309     & 11.65  & 3.942     & 13.06 & 4.444    & 12.22  & 4.722  	& 13.10    & 4.357 	\\
Ti I & 601.6995  & -3.630  & MFW	  &  3.52 & 4.51      & --     & --        &  1.70 & 4.429    &  1.85  & 4.818  	&  1.95    & 4.63  	\\
Ti I & 606.4626  & -1.944  & MFW          & --    & --        & --     & --        & 12.78 & 4.513    & 12.89  & 5.011  	& 14.13    & 4.665 	\\
Ti I & 609.1171  & -0.423  & MFW	  & 12.29 & 4.462     &  8.95  & 4.05      & 10.64 & 4.515    & 10.41  & 4.853  	& 11.87    & 4.662 	\\
Ti I & 609.2792  & -1.379  & MFW	  &  8.93 & 4.3       &  5.96  & 4.024     &  6.83 & 4.306    &  6.76  & 4.63   	&  8.42    & 4.568 	\\
%%%%%%%%%%%%%%%%
\\
\hline
\\
\multispan{3}{$\star$ Star showing TiO molecular bands in the spectra}
\\
\\
\multispan{10}{FMW -- Fuhr et al.~(1988)\hfill}\\
\multispan{10}{G03 -- Gratton et al.~(2003)\hfill}\\
\multispan{10}{SR -- Smith et al.~(1981)\hfill}\\
\multispan{10}{NBS -- Wiese et al.~(1969)\hfill}\\
\multispan{10}{MFW -- Martin et al.~(1988)\hfill}\\
\end{tabular}
}
\end{center}
\end{table*}

\begin{table}
\caption{Iron line list and adopted atomic parameters for star \#3800319. 
The measured equivalent width and the corresponding abundance obtained for each line are also reported}
\label{abund3800319}
\begin{center}
{\scriptsize
\begin{tabular}{rrrrrr}
\hline
\\
Ion  & $\lambda$ & log gf  & source of   & EW       &$\epsilon$\\
     & (nm)      &         & log gf      & (pm)     &          \\
     &           &         & (see notes) & 3800319  &          \\
\\
\hline
\\
%%%%%%%%%%%%%%%%%%%%%%%%%%%%%%%%%%%%%%%%%%%%%%%%%%%%%%%%%%%%%%%%%%%%%%%%%%%%
Fe I & 487.1318   &  -0.410  &  FMW   &   17.32   &  5.949 \\
Fe I & 491.8994   &  -0.370  &  FMW   &   17.84   &  5.958 \\
Fe I & 492.0502   &   0.060  &  FMW   &   30.56   &  6.368 \\
Fe I & 495.7298   &  -0.342  &  WBW2  &   17.69   &  5.878 \\
Fe I & 495.7596   &   0.127  &  WBW2  &   28.12   &  6.162 \\
Fe I & 500.6119   &  -0.615  &  FMW   &   22.38   &  6.571 \\
Fe I & 501.2068   &  -2.642  &  FMW   &   23.42   &  6.398 \\
Fe I & 505.1635   &  -2.795  &  FMW   &   22.54   &  6.528 \\
Fe I & 511.0413   &  -3.760  &  FMW   &   18.78   &  5.715 \\
Fe I & 517.1596   &  -1.793  &  FMW   &   20.64   &  6.045 \\
Fe I & 519.1454   &  -0.551  &  O     &   18.57   &  6.362 \\
Fe I & 519.2344   &  -0.421  &  O     &   14.89   &  5.643 \\
Fe I & 519.4941   &  -2.090  &  FMW   &   18.75   &  6.144 \\
Fe I & 522.7189   &  -1.228  &  O     &   23.52   &  5.867 \\
Fe I & 523.2940   &  -0.190  &  FMW   &   19.94   &  6.012 \\
Fe I & 526.6555   &  -0.490  &  FMW   &   19.96   &  6.379 \\
Fe I & 527.0356   &  -1.510  &  FMW   &   25.96   &  6.378 \\
Fe I & 532.4179   &  -0.240  &  FMW   &   17.49   &  6.123 \\
Fe I & 532.8039   &  -1.466  &  FMW   &   43.04   &  6.118 \\
Fe I & 532.8531   &  -1.850  &  O     &   22.24   &  6.302 \\
Fe I & 537.1489   &  -1.645  &  FMW   &   29.90   &  5.86  \\
Fe I & 539.7128   &  -1.993  &  FMW   &   26.15   &  5.884 \\
Fe I & 540.5775   &  -1.844  &  FMW   &   26.33   &  5.852 \\
Fe I & 542.9696   &  -1.879  &  FMW   &   32.89   &  6.209 \\
Fe I & 543.4524   &  -2.122  &  FMW   &   22.59   &  5.806 \\
Fe I & 544.6916   &  -1.930  &  FMW   &   28.62   &  6.095 \\
Fe I & 545.5609   &  -2.091  &  O     &   36.09   &  6.604 \\
Fe I & 561.5644   &  -0.140  &  FMW   &   19.44   &  6.333 \\
Fe I & 595.2718   &  -1.440  &  FMW   &    5.11   &  6.135 \\
Fe I & 602.7051   &  -1.210  &  FMW   &    6.22   &  6.195 \\
Fe I & 616.5360   &  -1.550  &  FMW   &    5.95   &  6.57  \\
Fe I & 670.3566   &  -3.160  &  FMW   &    4.69   &  6.127 \\
%%%%%%%%%%%%%%%%                            		
\\
\hline
\\
\multispan{6}{\scriptsize FMW -- Fuhr et al.~(1988)\hfill}\\
\multispan{6}{\scriptsize WBW2 -- Wolniket et al.~(1971)\hfill}\\
\multispan{6}{\scriptsize O -- O'Brian et al.~(1991)\hfill}\\
\end{tabular}
}
\end{center}
\end{table}

\end{document}